\documentclass[a4paper,fleqn,usenatbib]{mnras}

\usepackage{newtxtext,newtxmath}

\usepackage[T1]{fontenc}
\usepackage{ae,aecompl}

\usepackage{graphicx}	
\usepackage{amsmath}	
\usepackage{threeparttable}
\usepackage{booktabs}
\usepackage{wasysym} 

\newcommand{\A}{\text{\normalfont\AA}}

\def \ebmv{E(B-V)}
\def \ebmvs{ E_{s}{(B-V)} }
\def \ebmvn{ E_{n}{(B-V)} }
\def \halpha{H$\alpha$}
\def \Msol{{M}_{\odot}}

\def \logm{\log(M/\Msol)}
\def \lya{Ly$\alpha$}

\def \h2{{\rm H_{2}}}

\def \halpha{H$\alpha$}
\def \siiii{\ion{Si}{iii}}
\def \siiv{\ion{Si}{iv}}
\def \cii{[\ion{C}{ii}]}
\def \Cii{[\ion{C}{ii}]}

\def \oiicii{ \log(L_{\rm [OII]}/L_{\rm [CII]}) }
\def \oiiha{ \log(L_{\rm [OII]}/L_{\rm H\alpha}) }

\def \civ{\ion{C}{iv}}
\def \oii{[\ion{O}{ii}]}
\def \oiii{[\ion{O}{iii}]}
\def \nii{[\ion{N}{ii}]}
\def \neiii{[\ion{Ne}{iii}]}
\def \hii{\ion{H}{ii}}

\def \LIR{L_{{\rm IR}}}

\def \LCII{L_{{\rm CII}}}
\def \LOII{L_{{\rm OII}}}

\def \IRXB{IRX$-\beta$}
\def \dn4000{D_{{\rm n}}(4000) }

\def \iracA{[$3.6\,{\rm \mu m}$]$-$[$4.5\,{\rm \mu m}$]}

\def \thegalaxy{\textit{DC\_881725}}

\title[\oii$-$SFR relation and ISM at $z\sim4.5$]{The ALPINE-ALMA \Cii~Survey: Investigation of $10$ Galaxies at $z\sim4.5$ with \oii~and \Cii~Line Emission $-$ ISM Properties and \oii$-$SFR Relation}

\author[B. N. Vanderhoof et al.]{
Brittany N. Vanderhoof$^{1}$\thanks{E-mail: bxv3026@rit.edu},
A. L. Faisst$^{2}$,
L. Shen$^{3,4}$,
B. C. Lemaux$^{5}$,
M. B\'ethermin$^{6}$,\newauthor
\,\,P. L. Capak$^{7,8}$,
P. Cassata$^{9,10}$,
O. Le F\`evre$^{6}$,
D. Schaerer$^{11,12}$,
J. Silverman$^{13,14}$,
L. Yan$^{15}$,\newauthor
\,\,M. Boquien$^{16}$,
R. Gal$^{17}$,
J. Kartaltepe$^{1}$,
L. M. Lubin$^{5}$,
M. Dessauges-Zavadsky$^{11}$,\newauthor
\,\,Y. Fudamoto$^{18,19}$,
M. Ginolfi$^{20}$,
N. P. Hathi$^{21}$,
G. C. Jones$^{22,23}$,
A. M. Koekemoer$^{21}$,\newauthor
\,\,D. Narayanan$^{7,24,25}$,
M. Romano$^{9,10}$,
M. Talia$^{26,27}$, 
D. Vergani$^{26}$,
G. Zamorani$^{26}$
\\
$^{1}$ School of Physics and Astronomy, Rochester Institute of Technology, Rochester, NY 14623, USA\\
$^{2}$ IPAC, California Institute of Technology
1200 E California Boulevard, Pasadena, CA 91125, USA\\
$^{3}$ CAS Key Laboratory for Research in Galaxies and Cosmology, Department of Astronomy, University of Science and Technology of
China, Hefei 230026, China\\
$^{4}$ School of Astronomy and Space Sciences, University of Science and Technology of China, Hefei, 230026, China\\
$^{5}$ Gemini Observatory, NSF’s NOIRLab, 670 N. A’ohoku Place, Hilo, Hawai’i, 96720, USA \\
$^{6}$ Aix Marseille Universit\'e, CNRS, LAM (Laboratoire d'Astrophysique de Marseille) UMR 7326, F-13388, Marseille, France\\
$^{7}$ The Cosmic Dawn Center, University of Copenhagen, Vibenshuset, Lyngbyvej 2, DK-2100 Copenhagen, Denmark\\
$^{8}$ Niels Bohr Institute, University of Copenhagen, Jagtvej 128, 2200 Copenhagen, Denmark\\
$^{9}$ Dipartimento di Fisica e Astronomia, Universit\`a di Padova, vicolo dell'Osservatorio, 3 I-35122 Padova, Italy\\
$^{10}$ INAF, Osservatorio Astronomico di Padova, vicolo dell'Osservatorio 5, I-35122 Padova, Italy\\
$^{11}$ Observatoire de Gen\`eve, Universit\'e de Gen\`eve, 51 Ch. des Maillettes, 1290 Versoix, Switzerland\\
$^{12}$ Institut de Recherche en Astrophysique et Plan\'etologie—IRAP, CNRS, Universit\'e de Toulouse, UPS-OMP, 14, avenue E. Belin, F-31400 Toulouse, France\\
$^{13}$ Kavli Institute for the Physics and Mathematics of the Universe, The University of Tokyo, Kashiwa, 277-8583 (Kavli IPMU, WPI), Japan\\
$^{14}$ Department of Astronomy, School of Science, The University of Tokyo, 7-3-1 Hongo, Bunkyo, Tokyo 113-0033, Japan\\
$^{15}$ The Caltech Optical Observatories, California Institute of Technology, Pasadena, CA 91125, USA\\
$^{16}$ Centro de Astronom\'ia (CITEVA), Universidad de Antofagasta, Avenida Angamos 601, Antofagasta, Chile\\
$^{17}$ University of Hawai'i, Institute for Astronomy, 2680 Woodlawn Drive, Honolulu, HI 96822, USA\\
$^{18}$ Research Institute for Science and Engineering, Waseda University, 3-4-1 Okubo, Shinjuku, Tokyo 169-8555, Japan\\
$^{19}$ National Astronomical Observatory of Japan, 2-21-1, Osawa, Mitaka, Tokyo, Japan\\
$^{20}$ European Southern Observatory, Karl-Schwarzschild-Str. 2, D-85748, Garching, Germany\\
$^{21}$ Space Telescope Science Institute, 3700 San Martin Drive, Baltimore, MD 21218, USA\\
$^{22}$ Cavendish Laboratory, University of Cambridge, 19 J. J. Thomson Ave., Cambridge CB3 0HE, UK\\
$^{23}$ Kavli Institute for Cosmology, University of Cambridge, Madingley Road, Cambridge CB3 0HA, UK\\
$^{24}$ University of Florida Informatics Institute, 432 Newell Drive, CISE Bldg E251, Gainesville, FL 32611\\
$^{25}$ Department of Astronomy, University of Florida, 211 Bryant Space Sciences Center, Gainesville, FL 32611 USA\\
$^{26}$ INAF - Osservatorio di Astrofisica e Scienza dello Spazio di Bologna, via Gobetti 93/3, I-40129, Bologna, Italy\\
$^{27}$University of Bologna—Department of Physics and Astronomy "Augusto Righi" (DIFA), Via Gobetti 93/2, I-40129 Bologna, Italy\\ \vspace{-1.0cm}
}

\date{Accepted XXX. Received YYY; in original form ZZZ}

\pubyear{2021}

\begin{document}
\label{firstpage}
\pagerange{\pageref{firstpage}--\pageref{lastpage}}
\maketitle

\begin{abstract}
We present $10$ main-sequence \textit{ALPINE} galaxies ($\logm =  9.2-11.1$ and ${\rm SFR}=23-190\,{\rm M_{\odot}\,yr^{-1}}$) at $z\sim4.5$ with optical \oii~measurements from \textit{Keck}/MOSFIRE spectroscopy and \textit{Subaru}/MOIRCS narrow-band imaging. This is the largest such multi-wavelength sample at these redshifts, combining various measurements in the ultra-violet, optical, and far-infrared including \cii$_{158{\rm \mu m}}$~line emission and dust continuum from ALMA and H$\alpha$ emission from \textit{Spitzer} photometry.
For the first time, this unique sample allows us to analyse the relation between \oii~and total star-formation rate (SFR) and the interstellar medium (ISM) properties via \oii/\cii~and \oii/\halpha~luminosity ratios at $z\sim4.5$.
The \oii$-$SFR relation at $z\sim4.5$ cannot be described using standard local descriptions, but is consistent with a metal-dependent relation assuming metallicities around $50\%$ solar.
To explain the measured dust-corrected luminosity ratios of $\oiicii \sim 0.98^{+0.21}_{-0.22}$ and $\oiiha \sim -0.22^{+0.13}_{-0.15}$ for our sample, ionisation parameters $\log(U)< -2$ and electron densities $\log(\rm n_e / {\rm [cm^{-3}]}) \sim 2.5-3$ are required. The former is consistent with galaxies at $z\sim2-3$, however lower than at $z>6$. The latter may be slightly higher than expected given the galaxies' specific SFR.
The analysis of this pilot sample suggests that typical $\log(M/{\rm M_{\odot}}) > 9$ galaxies at $z\sim4.5$ to have broadly similar ISM properties as their descendants at $z\sim2$ and suggest a strong evolution of ISM properties since the Epoch of Reionisation at $z>6$.
\end{abstract}




\section{Introduction} \label{sec:intro}

Observing and understanding the interstellar medium (ISM) of galaxies is a key to understanding galaxy formation and evolution across cosmic time. Next to the far-infrared providing insights into the abundance of dust and gas, the rest-frame optical emission lines build an important basis to study the ISM. Specifically, these lines are sensitive to the instantaneous star formation, metal content of the gas, hydrogen densities, and ionisation rates among a variety of other key parameters.

During the recent years, substantial progress have been made in understanding the ISM properties of galaxies at $z=2-3$ through large spectroscopic survey such as the \textit{Keck Baryonic Structure Survey} \citep[KBSS,][]{STEIDEL14} or the \textit{MOSFIRE Deep Evolution Field} (MOSDEF) survey \citep[][]{KRIEK15}. 
Among others, one goal of such surveys is to pinpoint the properties of the ISM via three key parameters: the gas-phase metallicity, the electron density $n$ (in ${\rm cm^{-3}}$), and the dimensionless ionisation parameter $U$\footnote{The ionisation parameter is essentially the ratio of ionising photon density to hydrogen. It is also a measure for the radiation pressure feedback, as it reflects the radiation-to-gas-pressure ratio \citep[e.g.,][]{YEH12}.}. Through simultaneous analysis of multiple optical emission lines, it has been found that these parameters significantly evolve between $z=2$ and local galaxies, yet there is still some debate on the reasons for this change \citep[][]{STEIDEL14,STROM17,SANDERS16,MASTERS16}.
The evolution of ISM properties questions the reliability (or validity) of commonly used relations that have been calibrated to local galaxies. One of those is the relation between \oii~emission and star formation rate (SFR) as calibrated to local starburst galaxies \citep[][]{KENNICUTT98}. Similar to \halpha, \oii~is related to the current star formation in galaxies. However, as a forbidden line, the excitation of oxygen is sensitive to the abundances and ionisation state of the gas and not directly coupled to the ionisation radiation of young stars. The relation is therefore expected to change significantly as a function of these properties and hence likely redshift \citep[][]{KEWLEY04}. 

This motivates the interesting question of how the ISM changes at early cosmic times and whether common local relations are still valid at high redshifts. The latter is important to quantify, as more measurements of optical lines will be provided soon by the \textit{James Webb Space Telescope} (JWST). Of specific interest is the redshift range $z=4-6$. This epoch is fundamental to study the chemical and structural evolution of galaxies as it connects primordial galaxy formation during the Epoch of Reionisation at $z>6$ with mature galaxy evolution at the peak of cosmic SFR density at $z=2-3$. 
A large number of galaxies in this redshift range have been robustly identified via \lya~emission and ultra-violet (UV) absorption lines thanks to large spectroscopic surveys \citep[e.g.,][]{LEFEVRE15,HASINGER18} and are extensively studied in the UV.
Unfortunately, current facilities only allow the observation of one strong optical line (the \oii~doublet at rest-frame $3726\A$ and $3729\A$) at these redshifts, thus studies with multiple optical lines as the ones mentioned above are not possible.
So far, \oii~has only been observed in $4$ galaxies at $z=4-5$, with three of them lensed \citep[][]{SWINBANK07,SWINBANK09,TRONCOSO14,SHAPLEY17}. In \citet[][]{SHAPLEY17}, the \oii~as well as the detection of \neiii$_{\rm 3869}$~has been used to obtain the first direct measurement of the gas-phase metallicity at $z\sim4.5$ via optical lines. However, further determinations of the ISM properties of these galaxies have not been possible due to the lack of other emission lines.

In this paper, we investigate for the first time the ISM properties of high-$z$ galaxy population via the currently largest sample of $10$ typical main-sequence galaxies at $z\sim4.5$ with optical \oii~emission measurements. We mitigate the lack of optical emission lines by entering the far-infrared and in addition use spectral information from the rest-frame UV. Specifically, we are combining measurements of
\vspace{-0.1cm}
\begin{itemize}
    \item[\textit{(i)}] rest-UV absorption lines (obtained from \textit{Keck}/DEIMOS spectroscopy).
    \item[\textit{(ii)}] optical emission of \oii~(obtained from \textit{Subaru}/MOIRCS narrow-band imaging and \textit{Keck}/MOSFIRE spectroscopy) and \halpha~(obtained from \textit{Spitzer}/IRAC colours), and
    \item[\textit{(iii)}] singly ionised Carbon (\cii$_{158{\rm \mu m}}$~at $158\,{\rm \mu m}$, hereafter denoted as \cii) and $150\,{\rm \mu m}$ dust continuum from ALMA.
\end{itemize}

The galaxies, located in the \textit{COSMOS} field \citep[][]{SCOVILLE07}, are all part of the \textit{ALPINE} survey \citep[][]{ALPINE_BETHERMIN19,ALPINE_FAISST20,ALPINE_LEFEVRE20}, which currently provides the largest multi-wavelength post-reionisation dataset with observations from UV to far-infrared.
Thanks to the indirect measurement of metallicity from UV absorption lines as well as additional emission lines such as \halpha~and \cii, we can derive strong constraints on key ISM parameters such as metallicity, electron density, and ionisation parameter. Furthermore, the knowledge of the total SFR of the galaxies enables a first calibration of the relation between \oii~and SFR at $z\sim4.5$.

This paper is organised as follows: In Section~\ref{sec:sampledata} we present the spectroscopic and narrow-band observations of \oii, constraints on the metal content from rest-UV absorption lines, and the measurement of \halpha~emission from \textit{Spitzer}.
In the following section, we discuss the relation between total SFR and \oii~emission (Section~\ref{sec:discussionSFR}) and the \oii/\cii~and \oii/\halpha~luminosity ratio in the context of the ISM properties of our galaxies (Section~\ref{sec:discussioncloudy}). 
We conclude in Section~\ref{sec:end}.

Throughout this work, we assume a $\Lambda$CDM cosmology with $H_0 = 70\,{\rm km\,s^{-1}\,Mpc^{-1}}$, $\Omega_\Lambda = 0.70$, and $\Omega_{\rm m} = 0.30$. All magnitudes are given in the AB system \citep{OKE74} and stellar masses and SFRs are normalised to a \citet[][]{CHABRIER03} initial mass function (IMF) unless noted otherwise.

\section{Sample and Data} \label{sec:sampledata}

\begin{figure}
    \centering
    \includegraphics[width=0.48\textwidth]{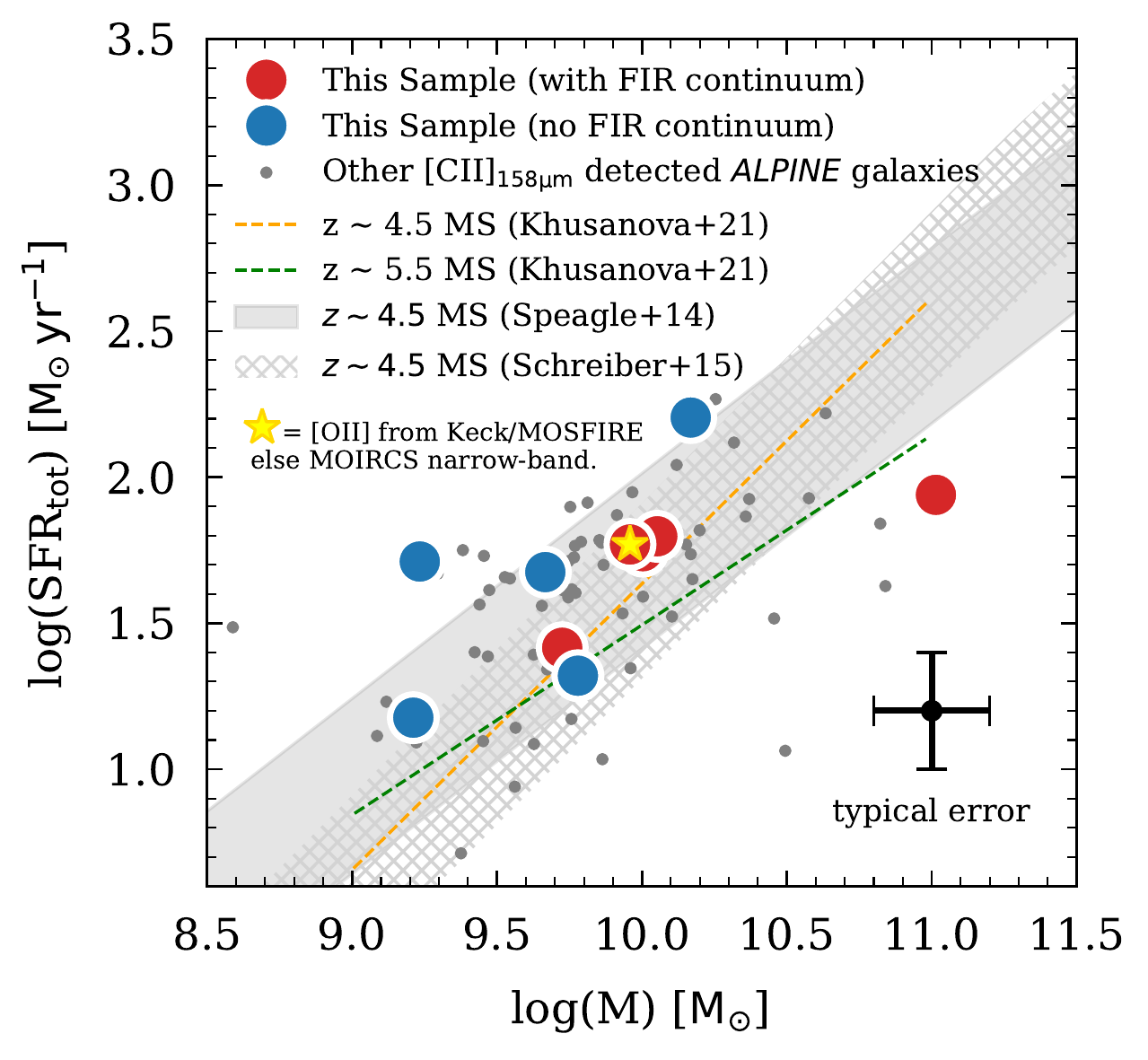}
    \vspace{-0.7cm}
    \caption{Relation between stellar mass and total SFR for the $10$ galaxies with \oii~and \cii~measurements studied in this work (red with far-infrared continuum detection; blue without). The total SFR in this case is computed by the UV$+$far-infrared continuum (if available) or \cii~alone according to the \citet[][]{ALPINE_SCHAERER20} relation (see Section~\ref{sec:discussionTotalSFR}). \thegalaxy, for which a spectroscopic measurement of \oii~exists, is shown as a yellow star. Other \textit{ALPINE} galaxies with \cii~measurements are shown in grey. We show the main-sequence (MS) parameterisations at $z\sim4.5$ from \citet[][]{SPEAGLE14} and \citet[][]{SCHREIBER15}, the latter extrapolated from \textit{Herschel} observations at $z\sim3.5$, as well as from \textit{ALPINE} at $z\sim4.5$ and $z\sim5.5$ \citep[][]{ALPINE_KHUSANOVA20}.
    }
    \label{fig:mainsequence}
    \vspace{-0.4cm}
\end{figure}

The $10$ galaxies at $z\sim4.5$ reside in the \textit{Cosmic Evolution Survey} \citep[COSMOS,][]{SCOVILLE07} field and are a subset of the \textit{ALPINE} survey \citep[][]{ALPINE_BETHERMIN19,ALPINE_FAISST20,ALPINE_LEFEVRE20}, which covers in total $118$ galaxies in COSMOS and the \textit{Extended Chandra Deep Field South} \citep[ECDFS,][]{GROGIN11,GIACCONI02}. The galaxies are covered by a wealth of observations at rest-frame UV (obtained by ground-based telescopes and the Hubble space telescope), optical (obtained by the \textit{Spitzer} space telescope), and far-infrared (obtained by ALMA) wavelengths.
The measurement of the $\sim150\,{\rm \mu m}$ dust continuum and \cii~line are detailed in \citet[][]{ALPINE_BETHERMIN19}.
Physical properties including stellar mass, SFR, dust attenuation (\ebmv), and UV continuum slope ($\beta$) are derived from the rest-frame UV to optical photometry by using the spectral energy distribution (SED) fitting code \texttt{LePhare} \citep[][]{ARNOUTS99,ILBERT06}. \citet[][]{BRUZUALCHARLOT03} composite stellar population models of different ages, metallicities, star-formation histories, and dust attenuation are assumed for the fitting as described in detail in \citet[][]{ALPINE_FAISST20}.
The galaxies are typical galaxies for their cosmic times and lie on the expected $z\sim4.5$ main-sequence (Figure~\ref{fig:mainsequence}). The total SFR in this case is computed from the UV and far-infrared continuum or alternatively from the \cii$-$SFR relation \citep[][]{ALPINE_SCHAERER20}, which was derived from the \IRXB~relation for galaxies without far-infrared continuum detection \citep[][]{ALPINE_FUDAMOTO20}.
The measurements coincide closely with two parameterisations of the main-sequence at $z\sim4.5$ from \citet[][]{SPEAGLE14} and \citet[][]{SCHREIBER15}. The latter is extrapolated from \textit{Herschel} observations at $z\sim3.5$.

\subsection{A Unique Sample of $z\sim4.5$ Galaxies with \oii~Measurements}\label{sec:oii_measurements}

In this study, we present the largest galaxy sample with \oii~measurement at $z\sim4.5$. For one of the $10$ galaxies, we obtain the \oii~measurement from spectroscopy while for the others \oii~is obtained from narrow-band photometry. In the following, we detail the measurements in both cases.

\begin{figure}
    \centering
    \includegraphics[width=0.48\textwidth]{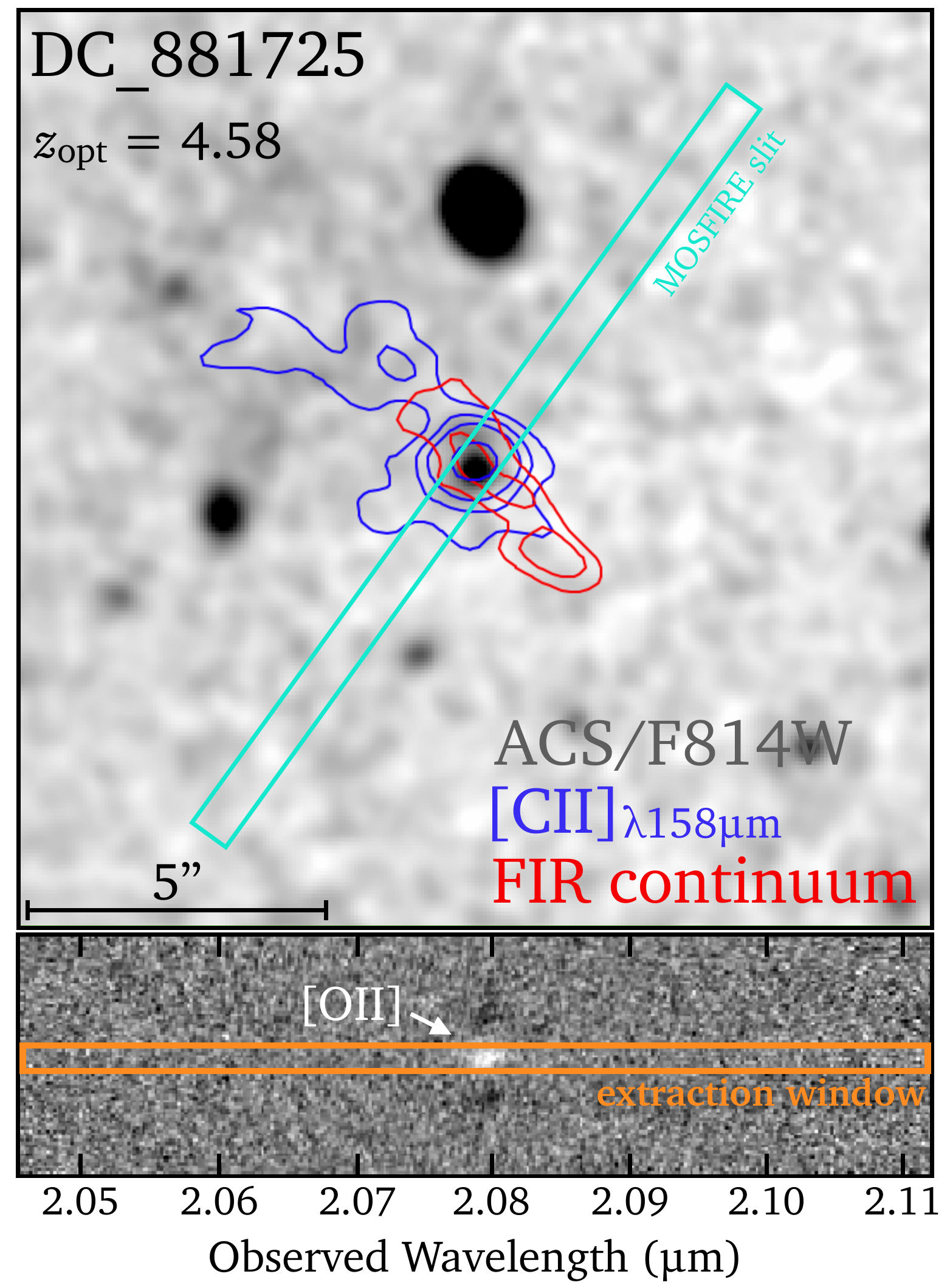}
    \vspace{-0.5cm}
    \caption{\textit{Keck}/MOSFIRE slit layout on \thegalaxy. \textit{Upper panel:} ACS/F814W image of \thegalaxy~overlaid with contours (1, 3, 5, and 10$\sigma$) of far-infrared continuum (red) and \Cii~emission (blue) observed by ALMA. The MOSFIRE slit is indicated as the cyan box. The F814W image \citep[][]{KOEKEMOER07} is smoothed by a Gaussian to represent the seeing (FWHM$\,=0.4\arcsec$) during the \textit{Keck} observations. \textit{Bottom panel:} Two-dimensional \textit{Keck}/MOSFIRE spectrum in K-band with \oii~emission indicated. The orange box visualises the extraction window ($6\,{\rm pixels}$ or $1.08\arcsec$).
    }
    \label{fig:slit}
    \vspace{-0.4cm}
\end{figure}

\subsubsection{\thegalaxy: Spectroscopic \oii~Measurement at $z=4.58$} \label{sec:keckdata}
\thegalaxy~ was observed at rest-frame optical wavelengths with the Multi-Object Spectrometer for Infrared Exploration \citep[\textit{MOSFIRE};][]{MCLEAN10,MCLEAN12} on the \textit{Keck} I telescope.
This galaxy was originally a filler target part of a \textit{MOSFIRE} program to observe a $z\sim2.2$ overdensity (PI: Nick Scoville). Hence the galaxy was basically randomly picked from the \textit{ALPINE} sample with the only requirement of having an \oii~flux \citep[estimated from the local SFR vs. \oii~relation;][]{KENNICUTT98} that is bright enough to be detected by the parent observations.
The observations cover the optical \oii~doublet at rest-frame wavelengths of $3726\,{\rm \A}$ and $3729\,{\rm \A}$. This is the first main-sequence galaxy at $z\sim4.5$ with spectroscopic \oii~measurement as well as observations of the far-infrared \cii~line emission and continuum at rest-frame $150\,{\rm \mu m}$.

The observations with \textit{MOSFIRE} were carried out on 2019 January 14 in $K$-band ($1.93 - 2.40\,{\rm \mu m}$) under clear sky conditions with an average seeing of 0.4$\arcsec$ and airmass of $1.1-1.2$. The galaxy was observed in MCSD mode with 16 reads, ABBA dither pattern, and a slit width of $0.7\arcsec$ for a total on-target exposure time of $1.2\,{\rm h}$ ($24\times 3\,{\rm min}$). The resolution of the spectrum corresponds to $R\sim 3600$, which would allow us theoretically to barely resolve the \oii~doublet if it were at higher signal-to-noise. Figure~\ref{fig:slit} shows \thegalaxy~in the Hubble ACS/F814W COSMOS mosaics \citep[][]{KOEKEMOER07} (convolved to match the resolution of the ground-based seeing at the time of the \textit{Keck} observations) with the \textit{MOSFIRE} slit overlaid. The far-infrared and \cii~emission observed with ALMA are superimposed as contours. The lower panel shows the two-dimensional spectrum and the window used to extract the one-dimensional spectrum.

The data was reduced and wavelength calibrated using the current MOSFIRE data reduction pipeline\footnote{\url{https://keck-datareductionpipelines.github.io/MosfireDRP/}}. Sky lines were used by the pipeline to perform a wavelength calibration.
The absolute flux calibration of the spectrum is detailed in Appendix~\ref{sec:absolutecalibration}. In brief, a magnitude $17$ standard star from the 2MASS star catalogue was included in the mask to derive a wavelength-dependent absolute flux calibration of the extracted one-dimensional spectrum.

\begin{figure}
    \centering
    \includegraphics[width=0.48\textwidth]{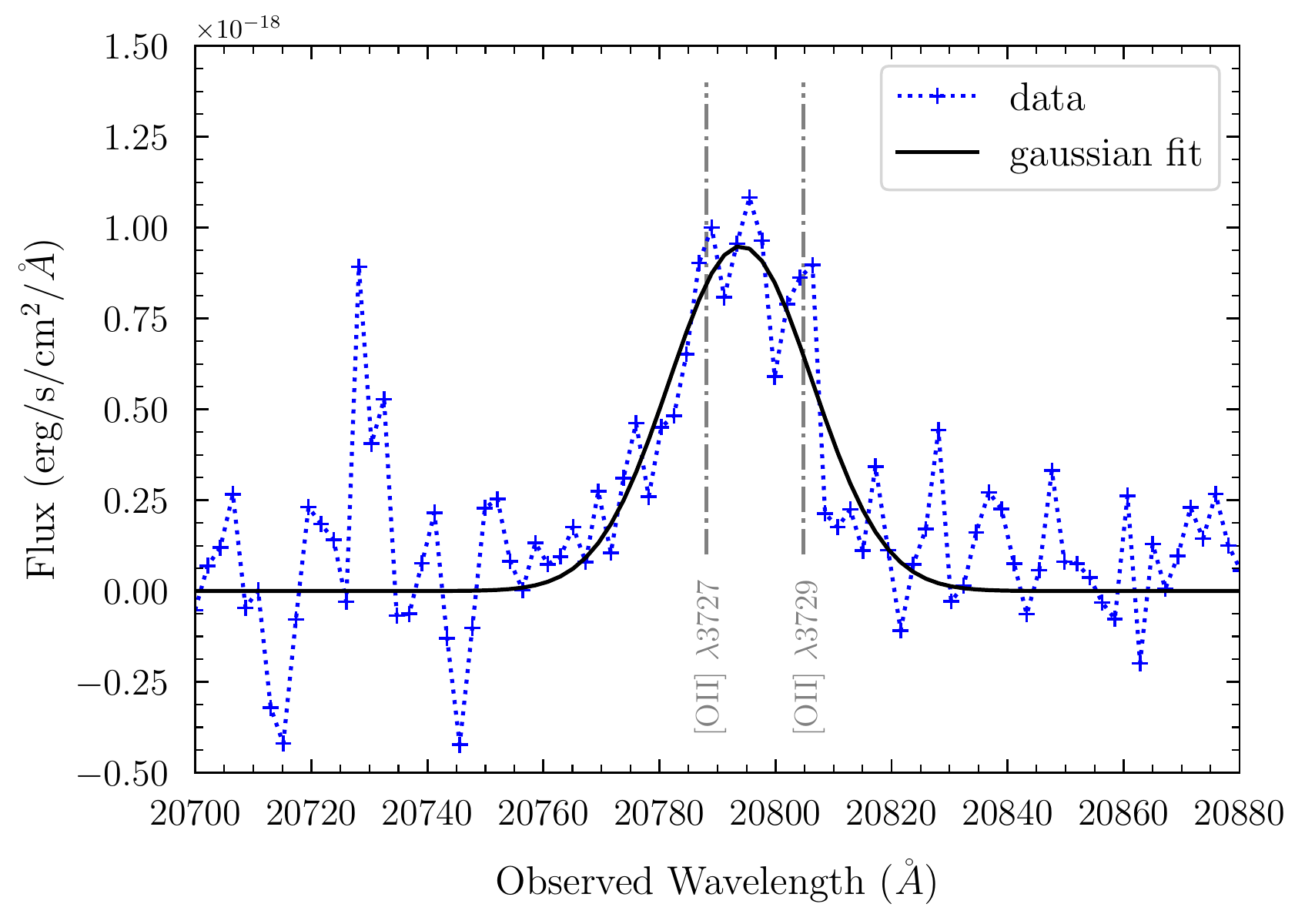}
    \vspace{-0.6cm}
    \caption{Extracted 1D MOSFIRE spectrum (blue) of \thegalaxy~at the location of the \oii~emission line doublet (indicated by the vertical dot-dashed lines). The Gaussian fit (assuming blended lines) is shown in black.}
    \label{fig:OIIgaussian}
    \vspace{-0.4cm}
\end{figure}

The one-dimensional spectrum of \thegalaxy~is extracted in a spatial window of $6$ pixels (corresponding to $1.08\arcsec$ at a pixel scale of $0.18\arcsec/{\rm px}$) around the centre of the identified \oii~emission line. The extracted spectrum in $\rm e^{-}\,s^{-1}$ is converted to $\rm erg\,s^{-1}\,cm^{-2}\,\A^{-1}$ using the wavelength dependent normalisation described in Appendix~\ref{sec:absolutecalibration}. The extracted spectrum is shown in Figure~\ref{fig:OIIgaussian} (blue dashed line) around the expected location of the \oii~doublet.

The \oii~line flux is derived by fitting a single Gaussian with variable mean, $\sigma$, and total flux to the one-dimensional spectrum (Figure~\ref{fig:OIIgaussian}, black solid line). We also tried to fit a double Gaussian to account for the doublet nature of the line. However, given the low signal-to-noise and resolution, no robust fit was obtained. The uncertainties of the total flux are determined using $500$ Monte Carlo samples for each of which the fluxes are changed according to their uncertainties (assumed Gaussian) obtained from the variance of the spectrum.
We find a total \oii~line flux of $3.00^{+0.47}_{-0.46}\times 10^{-17}\,{\rm erg\,s^{-1}\,cm^{-1}\,\A^{-1} }$  and measure an \oii~redshift of z$_{\oii}~=~4.5793$. The latter matches closely the redshift determined from \cii~($\sim123\pm30\,{\rm km\,s^{-1}}$ blue-shifted w.r.t. \oii) but is significantly red-shifted with respect to Ly$\alpha$ by $328\pm30\,{\rm km\,s^{-1}}$.\footnote{The velocity shifts quoted assume equal flux of the two blended \oii~lines. We estimate an uncertainty of $\pm30\,{\rm km\,s^{-1}}$ assuming \oii$_{\rm \lambda 3729}$/\oii$_{\rm \lambda 3726}$ ratios between $0.35$ and $1.5$, corresponding to the high and low electron density regime \citep[][]{OSTERBROCK74}.} These quantities as well as the \oii~luminosity, L$_{\oii}$, are listed in Table~\ref{tab:keckproperties} along with other properties measured from the ancillary data.

\subsubsection{\oii~Measurements from Narrow-Band Imaging}\label{sec:narrowbanddata}


We complement \thegalaxy~with $9$ additional galaxies at $z\sim4.5$ which have \oii~luminosities derived from narrow-band imaging taken with the Multi-Object Infrared Camera and Spectrograph \citep[MOIRCS;][]{Ichikawa2006,Suzuki2008} on the \textit{Subaru} Telescope. The observations were part of the \textit{Charting Cluster Construction with VUDS} \citep{LEFEVRE15} \textit{and ORELSE} \citep{Lubin2009} survey \citep[C3VO,][]{Lemaux20,Shen21}. 
In brief, the $9$ \textit{ALPINE} galaxies fall in the footprint of three pointings targeting the massive proto-cluster \textit{PCl J1001+0220} at $z\sim4.57$ \citep{Lemaux2018} and its surrounding in the COSMOS field. Similar to \thegalaxy, they are therefore randomly selected from the \textit{ALPINE} sample but without imposed restrictions on the \oii~flux. The $9$ galaxies may be part of the proto-cluster structure, however, as shown by studies at lower redshifts \citep[e.g.,][]{DARVISH15}, we do not expect significant differences in their properties compared to field galaxies.

The observations were executed between February 2 2020 and January 31 2021, using NB2071 ($2.043-2.097\,{\rm \mu m}$) and NB2083 ($2.056-2.110\,{\rm \mu m}$) narrow-band filters, under seeing that ranged from $\sim0.4-0.8\arcsec$ and conditions that varied from light cirrus to photometric. The total effective integration time across all three pointings and two filters was approximately $18$ hours, which was split into $150{\rm s}$ individual exposures. A standard circular dither pattern with a set of 10 dithers was adopted. 

All raw data were reduced with the IRAF based reduction pipeline \textsc{MCSRED2} \citep{Tanaka2011}, which performed flat-fielding, masking objects, and sky subtraction. Astrometry calibration was performed by \textsc{SCAMP} \citep{Bertin2006} for every individual exposure. For each filter and chip, a final narrow-band image was stacked using \textsc{SWarp} \citep{Bertin2002}. 
The $5\sigma$ limiting magnitudes of the images have a median of $23.1\,{\rm mag}$ and in a range of $22.7-23.3\,{\rm mag}$, where $\sigma$ is measured from the flux scatter of randomly distributed $2\arcsec$ aperture. 
Note that some exposures were shallower due to the filter wheel occultation. Finally, photometric calibration was performed to the \textit{Ks}$-$band image from the UltraVISTA DR4 \citep{McCracken2012} by selecting a set of bright but unsaturated point sources as the reference stars.

For the source detection and extraction on each narrow-band image, \textsc{SExtractor} \citep{Bertin1996} was used in dual-image mode using the much deeper ($24.8\,{\rm mag}$ at $5\sigma$) \textit{Ks}$-$band image as the detection image.\footnote{We set detection parameters of \texttt{DETECT\_MINAREA}~$=5$, \texttt{DETECT\_THRESH}~$=1.5$, and \texttt{ANALYSIS\_THRESH}~$=1.5$. The latter two are relative ($\sigma$) thresholds.} 
The point-spread functions (PSFs) of the narrow-band images were degraded to match the \textit{Ks}-band image. 
In details, the PSFs for the narrow-band and \textit{Ks}-image were derived using the \textsc{PSFex} code \citep{Bertin2011}. Subsequently, the \textsc{Photutil} package with a Split Cosine Bell window was used to generate a matching kernel between two PSFs. The uncertainties on the fluxes estimated by \textsc{SExtractor} were scaled to account for correlated pixel noise following the method of \citet{pelliccia2021}.

The following system of two equations involving the narrow-band and the underlying \textit{Ks}-band photometry was solved for the continuum ($f^{\lambda}_{\rm cont}$) and \oii~line flux ($f^{\lambda}_{\rm line}$) at wavelength $\lambda$ of the \oii~line \citep[c.f.][]{Hu2019}:
\begin{equation} \label{eq_oii}
\bar{f}^{\rm \lambda}_{\rm NB/Ks} = \frac{\int\left(f^{\lambda}_{\rm line} + f^{\lambda}_{\rm cont}\right) T^{\lambda}_{\rm NB/Ks} d\lambda}{\int T^{\lambda}_{\rm NB/Ks} d\lambda}.
\end{equation}

\noindent In the above equation, $\bar{f}^{\lambda}_{\rm NB}$ and $\bar{f}^{\lambda}_{\rm Ks}$ denote the detected flux densities in each narrow-band and the \textit{Ks}-band, and $T^{\lambda}_{\rm NB}$ and $T^{\lambda}_{\rm Ks}$ are the corresponding filter transmission functions.
In the calculation, we assumed an \oii~line profile resembling a $\delta$-function in each narrow-band filter and a flat continuum profile. 
As the observed line width of the \oii~doublet ($29.01_{-4.64}^{+5.33}$\AA, see Figure~\ref{fig:OIIgaussian}) is much narrower than the width of the narrow-band filter ($\sim 270\,{\rm \A}$), so the \oii~line profile does not affect the result of this calculation.
The uncertainties of the \oii~fluxes were obtained using Monte-Carlo iterations for each of which the narrow-band and \textit{Ks}-band fluxes were drawn from a Gaussian distribution with a width corresponding to their associated uncertainties. The \oii~flux uncertainty is then defined as the 16$^{\rm th}$ and 84$^{\rm th}$ quantiles of the resulting \oii~flux distribution.
Finally, an inverse-variance weighting scheme was adopted to combine fluxes obtained from the two narrow-band filters, with weights defined as $W_i = T_i^2 / \sigma_i^2$, where $T_i$ is the narrow-band transmission curve at the \cii~redshift and $\sigma_i$ is the narrow-band flux uncertainty. 

Table~\ref{tab:narrowbandproperties} lists the measured \oii~luminosities as well as other properties derived from the ancillary data for these $9$ galaxies.

\subsection{Estimate of Metal-Enrichment from Rest-Frame UV Absorption Lines}
\label{sec:metallicity}

Commonly, the gas-phase metal content of galaxies is measured by flux ratios of bright optical lines such as \oii, \oiii, and \halpha~\citep[e.g.,][]{PETTINI04,MAIOLINO19}. Recently, thanks to ALMA, also far-infrared emission lines such as \nii$_{205{\rm \mu m}}$, \oii$_{88{\rm \mu m}}$, and \cii~are used to estimate the gas-phase metal content of high-redshift galaxies \citep[e.g.,][]{VALLINI21,YANGLIDZ20,JONESTUCKER20,PAVESI19,CROXALL17}.
The former are not accessible at these redshifts currently (except \oii), while the latter are not available for our galaxies as their observations are costly (except \cii) or only possible for galaxies at $z>6$ (in the case of \oii$_{88{\rm \mu m}}$).

Nonetheless, we can estimate the metal content of our galaxies from the exquisite rest-frame UV spectra in an indirect way. Specifically, the strength, i.e., equivalent-width (EW; note that EWs of absorption lines are defined positive), of rest-frame UV absorption lines correlates with metal content \citep[e.g.,][]{LEITHERER11}. This correlation has different origins including the evolutionary stage of stellar populations, metal-dependent winds impacting the width of absorption lines, or dust extinction. The technique has already been applied to similar main-sequence galaxies at $z\sim4-6$ \citep[][]{FAISST16b,ANDO07} and has been tested empirically at lower redshifts \citep[][]{FAISST16b}.
The S/N of individual rest-UV spectra of our galaxies is too low to estimate their metal content. We therefore perform median stacking the $10$ spectra before we compute the EWs of several absorption complexes around \siiii~($\sim 1300\,{\rm \A}$), \ion{C}{ii}~($\sim 1335\,{\rm \A}$), \siiv~($\sim 1400\,{\rm \A}$), and \civ~($\sim 1550\,{\rm \A}$). The uncertainties are computed via Monte-Carlo iterations including the spectral variances. We then use the relations between absorption line EW and metallicity as calibrated via galaxies in the local universe as well as $z\sim2$ and $z\sim3$ in \citet[][]{FAISST16b}.

\begin{figure}
    \centering
    \includegraphics[width=0.48\textwidth]{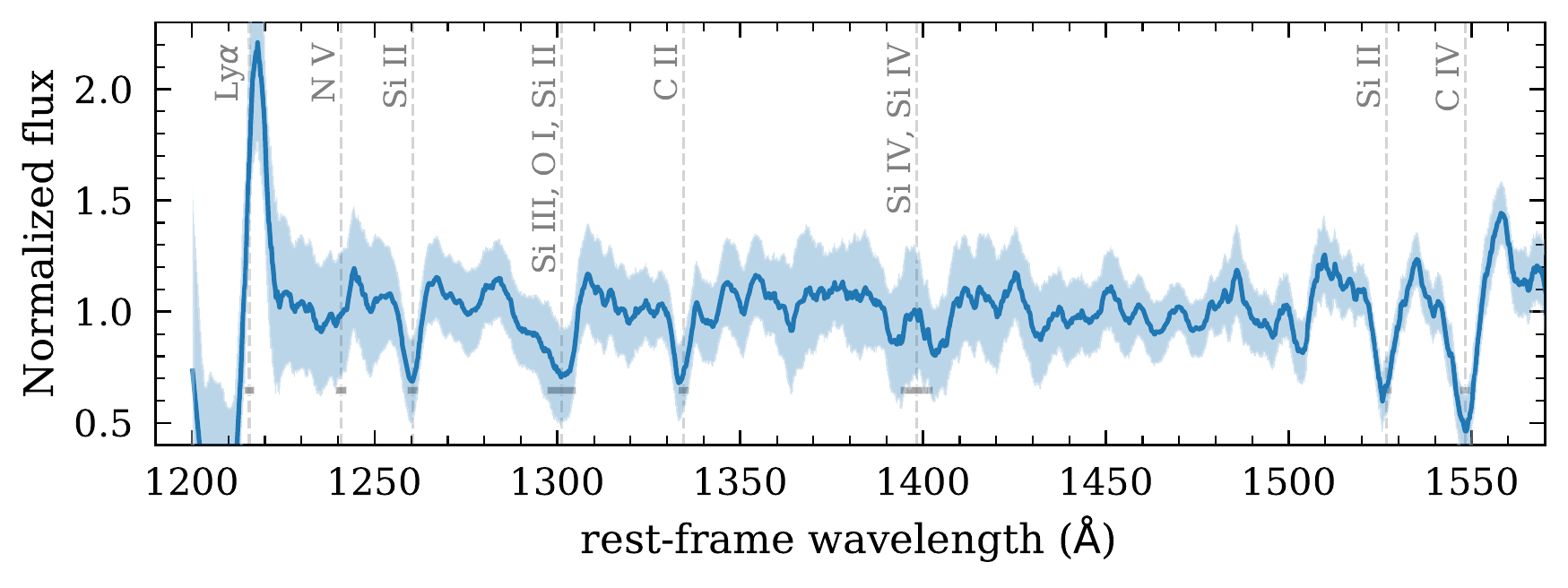}\\
    \includegraphics[width=0.48\textwidth]{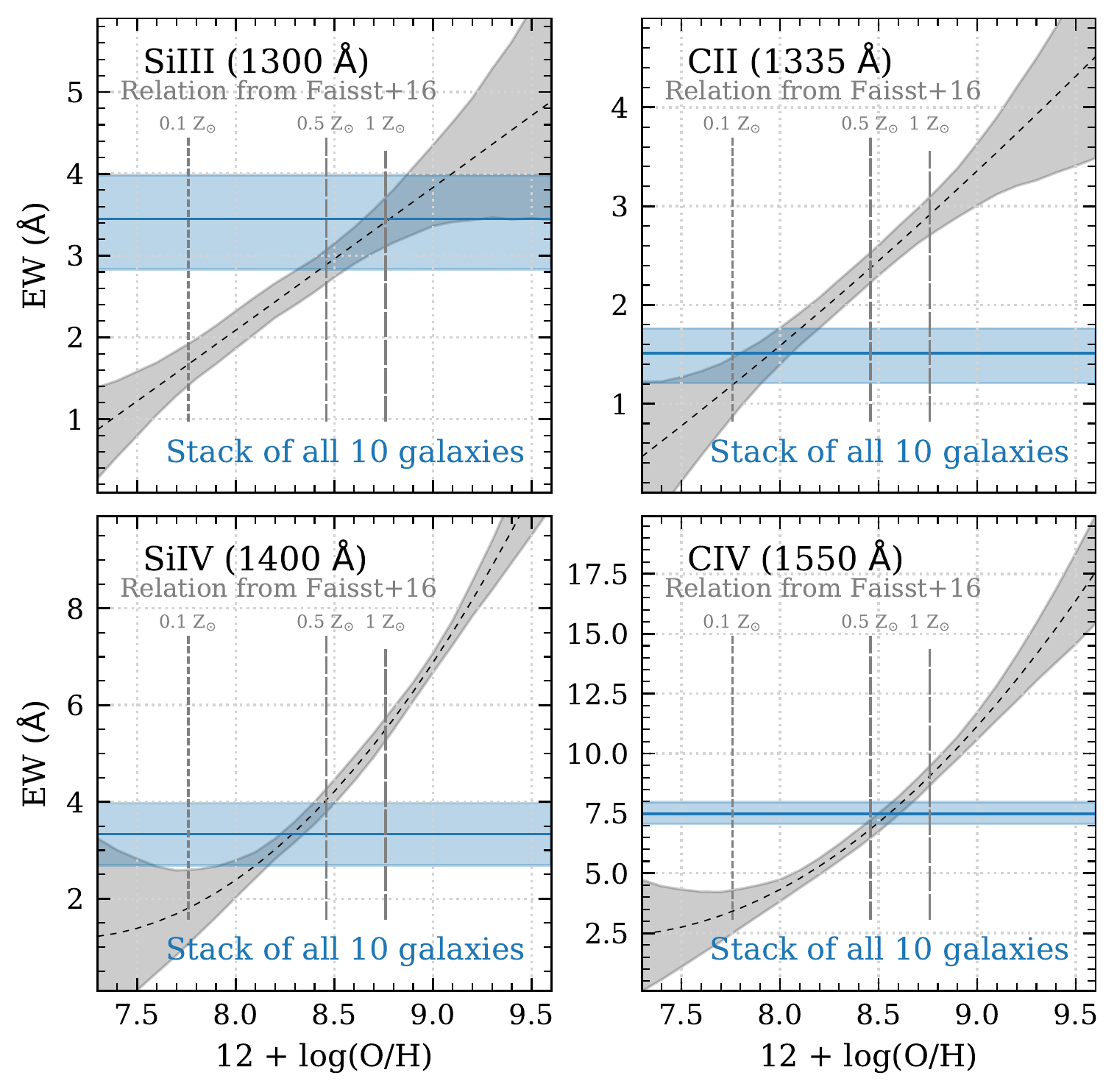}
    \caption{\textit{Top panel:} Flux-normalized stacked UV spectrum of the $10$ galaxies studied in this work. Prominent spectral features are indicated. The light area shows the $1\sigma$ scatter determined from individual variances of the spectra. \textit{Bottom panels:} The EWs of rest-frame UV absorption lines provide an indirect measure of the metal content. The blue swaths show the measured EWs of each absorption complex from a stack of all $10$ galaxies in our sample. The grey swath shows the calibration between EW and metallicity from \citet[][]{FAISST16b}. The vertical dashed lines mark $0.1$, $0.5$, and $1\,{\rm Z_{\odot}}$.
    }
    \label{fig:metallicity}
\end{figure}

The stacked spectrum and the results from this analysis are shown in Figure~\ref{fig:metallicity}. The blue horizontal swaths show the measured EWs of each absorption complex with uncertainty and the grey swaths show the calibration derived in \citet[][]{FAISST16b}.
Looking at the individual absorption complexes suggests average metallicites between $10\%$ and $50\%$ of solar, except for \siiii, which is consistent with solar metallicity.
We note that even for the stack the S/N is relatively low, which is likely the cause for the apparent differences between the individual absorption complexes.
Combining the probability distribution functions of all absorption complexes, we find an average metallicity of $12+\log{(\rm O/H}) = 8.41^{+0.31}_{-0.54}$. This corresponds to roughly $\sim50\%$ of the solar metallicity\footnote{We assume a solar oxygen abundance of ${\rm \log(O/H)+12} = 8.76$ according to the ``galactic concordance abundances'' \citep[][]{NIEVA12,NICHOLLS17}, which is close to the primordial solar abundance \citep[][]{ASPLUND09}.}.
Such values are consistent with the expected metallicity based on the stellar masses of our galaxies at these redshifts \citep[][]{FAISST16b,ANDO07}.

\subsection{Photometric Measurement of \halpha~from \textit{Spitzer}}\label{sec:ha_measurements}

The coverage of our galaxies by \textit{Spitzer} allows us to photometrically estimate their \halpha~emissions via the \iracA~colours. 
This method has been successfully used by several studies very early on \citep[e.g.,][]{SHIM11,STARK13,MARMOLQUERALTO16,RASAPPU16} and it is shown to result in photometric \halpha~measurements that are statistically consistent with spectroscopy \citep[e.g.,][]{FAISST16a}.
The detailed derivation of this measurement for the \textit{ALPINE} galaxies is described in \citet[][]{ALPINE_FAISST20}. The luminosities are corrected for dust in the same way as \oii.
As an additional check, we compare our \halpha~measurement to the \textit{EL-COSMOS} catalogue \citep[][]{SAITO20}, which contains \halpha~emission line predictions from SED fitting for all galaxies from the \citet[][]{LAIGLE16} COSMOS catalogue. We find good agreement within a factor of $2-5$ (see Appendix~\ref{sec:elcosmos}). We find a similar agreement when comparing the \oii~luminosities provided in the \textit{EL-COSMOS} catalogue to our spectroscopic and narrow-band imaging measurements, which verifies the reliability of the catalogue.

\section{Discussion}

\begin{figure*}
    \centering
    \includegraphics[width=0.51\textwidth]{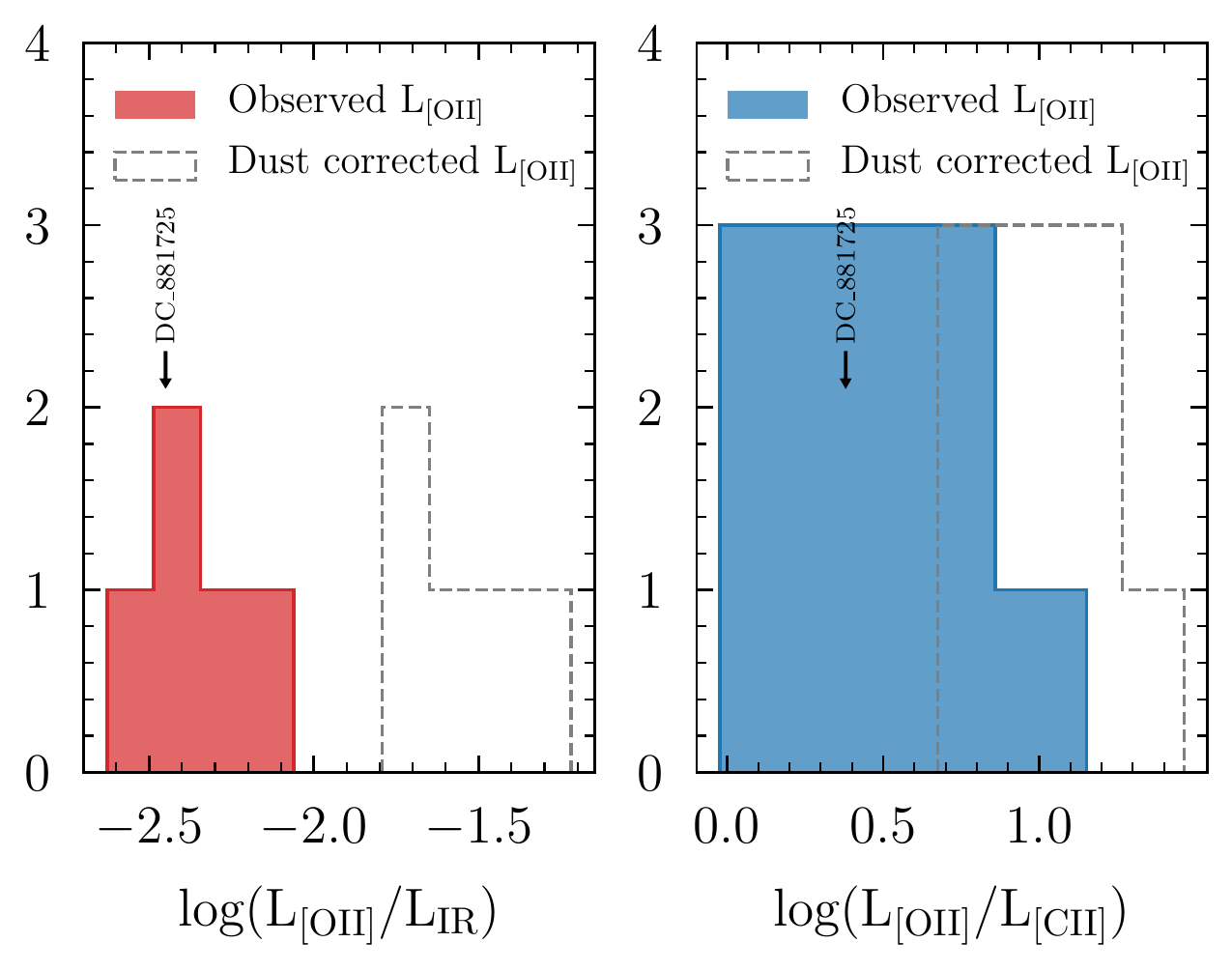}
    \includegraphics[width=0.45\textwidth]{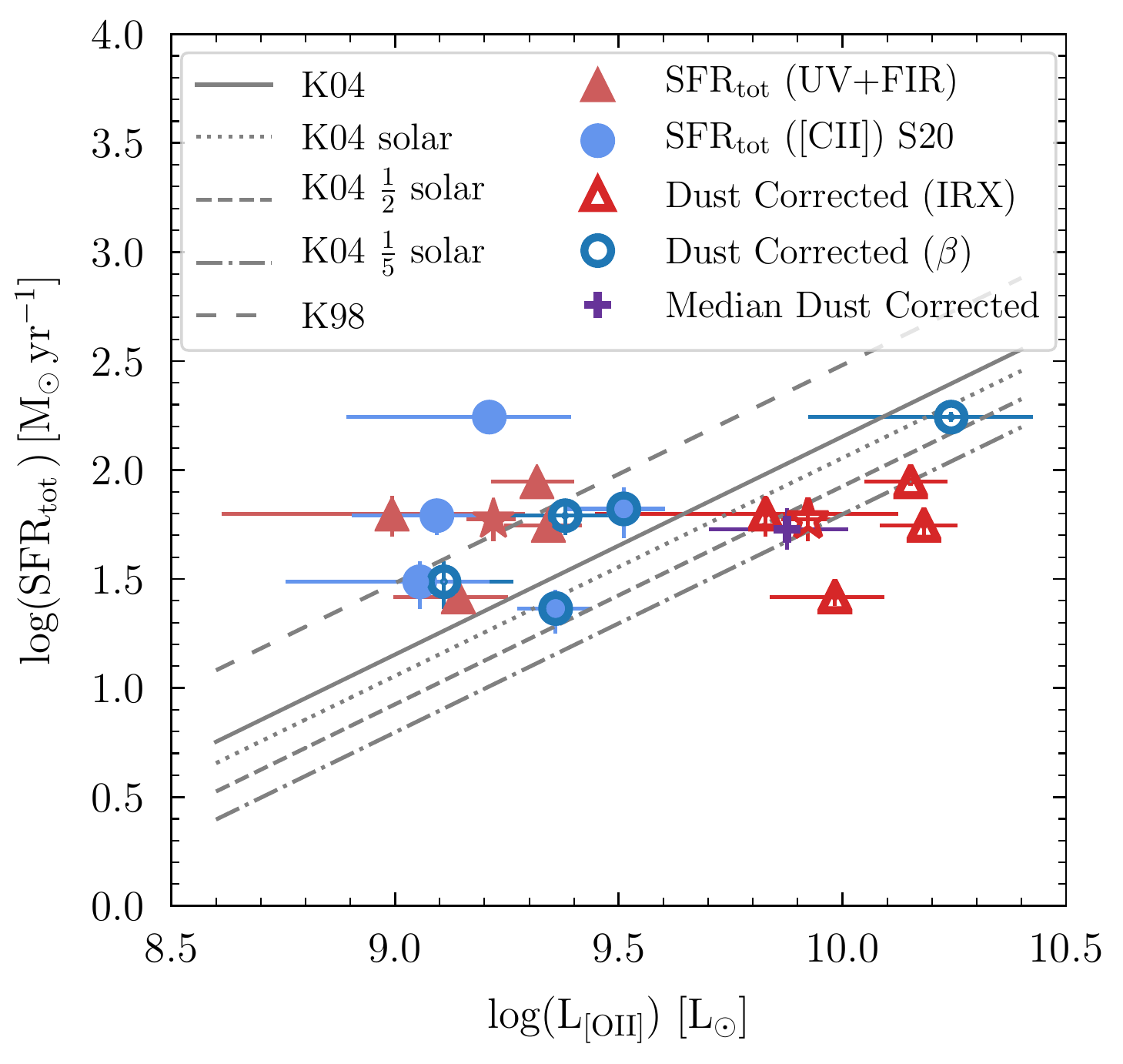}
    \vspace{-0.3cm}
    \caption{ \textit{Left panels:} Histograms of the $\LOII/\LIR$ (left) and $\LOII/\LCII$ (right) ratios for the $10$ galaxies. The observed ratios for \thegalaxy~are indicated with arrows. The filled histograms show observed ratios (not corrected for dust) and the dashed histograms show dust-corrected ratios (see text for details).
    \textit{Right panel:} Relation between \oii~emission and total SFR. The red (blue) symbols show galaxies with (without) far-infrared detections from ALMA. In the former case, the total SFR is ${\rm SFR_{UV+IR}}$; in the latter case, the total SFR is derived from \cii~emission. The filled lighter symbols show observed \oii~luminosities (not corrected for dust) while the empty darker symbols show dust corrected \oii~luminosities. For far-infrared detected galaxies, the dust correction is derived from the IRX value directly. For the others, the relation between $\beta$ and IRX is used (see Section~\ref{sec:discussionOIISFR}). Note that some galaxies have little dust, in which cases the filled and empty symbols coincide.
    \thegalaxy~is shown as star. The purple ``plus'' denotes the median dust corrected \oii~luminosity.
    We also show different empirical \oii$-$SFR relations from \citet[][long dashed line]{KENNICUTT98} and the updated version from \citet[][solid line]{KEWLEY04}, as well as the metallicity dependent models from \citet[][dotted, short dashed, and dot-dashed lines for solar, $50\%$ solar, and $20\%$ solar metallicity]{KEWLEY04}.
    }
    \label{fig:sfr_oii}
    \vspace{-0.3cm}
\end{figure*}

\subsection{The \oii$-$SFR relation at $z\sim4.5$}\label{sec:discussionSFR}

We first investigate whether the local relation between \oii~emission and total star formation is still valid at $z\sim4.5$. For this we make use of our multi-wavelength measurements of the total SFR from UV and far-infrared data.

\subsubsection{Total SFRs Derived from UV$+$Far-Infrared Continuum and \cii~Emission}\label{sec:discussionTotalSFR}

In order to relate \oii~emission to total star formation, we have to provide measurements of the latter that are as robust as possible. Our sample allows us to do this via the combination of measurement of UV and far-infrared continuum as well as \cii~emission.

For galaxies detected in far-infrared continuum ($5$ out of $10$), we compute the total SFR as the sum of UV and dust obscured star formation. Specifically, we make use of the relations detailed in \citet[][]{KENNICUTT98} \citep[see also][]{KENNICUTT21},
\begin{equation}\label{eq:uv}
    {\rm SFR_{UV}\,(M_{\odot}\,yr^{-1})} = 0.79 \times 10^{-28} L_{\rm \nu} ({\rm erg\,s^{-1}\,Hz^{-1}})
\end{equation}
and
 \begin{equation}\label{eq:fir}
    {\rm SFR_{IR}\,(M_{\odot}\,yr^{-1})} = 2.54 \times 10^{-44} L_{\rm IR} ({\rm erg\,s^{-1}}).
\end{equation}
These SFR values have already been converted from a \citet[][]{SALPETER55} to a \citet[][]{CHABRIER03} IMF by division of a factor of $1.77$ (in linear scale). For details on the derivation of the UV and far-infrared luminosity we refer to \citet[][]{ALPINE_FAISST20} and \citet[][]{ALPINE_BETHERMIN19}, respectively. Note that the derivation of $L_{\rm IR}$ depends strongly on the shape of the far-infrared SED, with specifically the dust temperature playing an important role \citep[e.g.,][]{FAISST17b}.
Here, we make use of the stacked far-infrared SED derived from Herschel, SCUBA, and ALMA photometry of $z=4-6$ galaxies in COSMOS with similar SFR and stellar masses as the \textit{ALPINE} galaxies \citep[see][]{ALPINE_BETHERMIN19}. The dust temperature of this stack is $43\pm5\,{\rm K}$, which is consistent with the individual dust temperatures measured for $4$ \textit{ALPINE} galaxies with sufficient far-infrared coverage from ALMA \citep[][]{FAISST20b}.

For the remaining five galaxies without far-infrared continuum detection, we use the relation between \cii~line emission and total SFR as presented by \citet[][]{ALPINE_SCHAERER20},
 \begin{equation}\label{eq:cii}
    \log({\rm SFR_{[CII]}/ [M_{\odot}\,yr^{-1}]}) = \frac{\log({\rm L_{[CII]} / L_{\odot}}) - 6.61}{1.17}.
\end{equation}
The above relation is based on total SFRs of all \textit{ALPINE} galaxies (with limits from \cii~non-detection properly taken into account). For galaxies without far-infrared detection, the \IRXB~relation \citep[][]{ALPINE_FUDAMOTO20} is used to derive their far-infrared luminosities and total SFRs.  

The total SFR for the $9$ narrow-band detected galaxies are listed in Table~\ref{tab:narrowbandproperties}. The total SFRs of \thegalaxy~derived from UV and far-infrared as well as \cii~for comparison are listed in Table~\ref{tab:sfrs}. Note that the \cii~and far-infrared (if available) derived SFRs are very comparable as the former have been calibrated by the latter in \citet[][]{ALPINE_SCHAERER20}.

We can also compare different \cii$-$SFR relations derived in \citet[][]{DELOOZE14} with our total SFRs derived for the $5$ galaxies with far-infrared continuum detection.
These relations have been calibrated using low-metallicity dwarfs ($12+\log({\rm O/H}) \sim 7.1-8.4$), starbursts (and \hii~regions), galaxies with an active galactic nucleus (AGN), ultra-luminous infrared galaxies (ULIRGs), and high-redshift galaxies (mostly at $z\sim1-3$).
We find that \textit{(i)} the \cii$-$SFR relation of local metal-poor dwarfs underestimates the total SFR by $0.43\pm0.13\,{\rm dex}$, while \textit{(ii)} the relation derived from ULIRGs overestimates the total SFRs by $0.77\pm0.15\,{\rm dex}$. The other calibrations are roughly consistent with our derived values within $1\sigma$ uncertainties.
Similar results are found for the remaining $5$ galaxies with SFRs derived from \cii~using the \citet[][]{ALPINE_SCHAERER20} relation.
This SFR comparison suggests that our galaxies are not comparable to either local low-metallicity dwarfs or local ULIRGs. This is not  surprising as rest-frame UV absorption line spectroscopy suggests that our galaxies are rather metal enriched (see Section~\ref{sec:metallicity}). Furthermore, their far-infrared luminosities are about an order of magnitude lower than those of local ULIRGs included in \citet[][]{DELOOZE14} ($L_{\rm IR} \sim 3\times 10^{12}\,{\rm L_{\odot}}$).
Table~\ref{tab:sfrs} shows a detailed list of SFRs derived by the different \cii$-$SFR relations for the example of \thegalaxy.

\subsubsection{The \oii$-$SFR relation at $z\sim4.5$}\label{sec:discussionOIISFR}

With measured robust total SFRs for our galaxies, we can now relate these values to the \oii~emission.

The left panels of Figure~\ref{fig:sfr_oii} show the histograms of the $\LOII/\LIR$ and $\LOII/\LCII$ luminosity ratios. We find observed median ratios of $\oiicii~=~0.39^{+0.35}_{-0.32}$ and $\log(L_{\rm [OII]}/L_{\rm IR})=-2.45^{+0.23}_{-0.09}$. The corresponding dust-corrected ratios are $\oiicii~=~0.98^{+0.21}_{-0.22}$ and $\log(L_{\rm [OII]}/L_{\rm IR})=~-1.64^{+0.23}_{-0.13}$, respectively.

With \oii~being an optical line, it has to be corrected for the effect of dust attenuation.
We compute \oii~dust correction factors using the stellar continuum dust attenuation $\ebmvs$ values via
\begin{equation}
    f_{\rm corr} = 10^{0.4 \ebmvs k_{\lambda}/f},
\end{equation}
where $k_{\rm \lambda}$ is the reddening curve with $\lambda=3727\,{\rm \A}$ and we assume a differential dust attenuation factor between stellar continuum and nebular emission of $f=0.44$ \citep[][]{CALZETTI00}\footnote{The ``$f-$factor'' is the differential dust attenuation between nebular emission and stellar continuum and is defined as $f = {E_{s}(B-V)}/{E_{n}(B-V)}$.}. 
The $\ebmvs$ values are estimated by the following procedure. For galaxies that are detected in far-infrared continuum, we used their IRX ratio ($\equiv \log(L_{\rm IR}/L_{\rm UV})$) to estimate the dust attenuation at rest-frame $1600\,{\rm \A}$ ($A_{\rm 1600}$) via the relation given in \citet[][]{HAO11}\footnote{Note that this parameterization is consistent with earlier derivations \citep[e.g., from][]{BUAT05,BURGARELLA05,KONG04,MEURER99}.},
\begin{equation}\label{eq:a1600}
    A_{\rm 1600} = 2.5 \log(1 + 0.46 \times 10^{\rm IRX}).
\end{equation}
The nebular $\ebmv$~values, $\ebmvn$, are then estimated using the relation $A_{\rm 1600} = 4.39 \times \ebmvn$ \citep[][]{CALZETTI00}. Other parameterisations \citep[e.g.,][]{REDDY15} lead to values that are different by less than $20\%$. 
For galaxies that are not detected in far-infrared continuum, we derive the stellar $\ebmvs$~values from the UV continuum photometry. Specifically, we explore the values from SED fitting directly as well as derived from the UV continuum slope ($\beta$) using the dust parameterisations by \citet[][]{CALZETTI00} and \citet[][]{REDDY15}. As an additional comparison, we also derive IRX values directly from the $\beta$ slopes using the \IRXB~relation fit for \textit{ALPINE} galaxies at $z\sim4.5$ in \citet[][]{ALPINE_FUDAMOTO20}.
We find that all four methods agree within $0.05\,{\rm mag}$, which results in $<30\%$ differences in the dust factors for \oii.
All in all, we find \oii~dust attenuation factors (linear) between $1-5$, except in one case where we find a value of $f_{\rm corr}\sim10$. The different $k_{\rm \lambda}$ parameterisations used above result in variations of $<40\%$ (leading to uncertainties of $<0.14\,{\rm dex}$ in \oii~luminosity).
The differential dust attenuation between nebular emission and stellar continuum is by far the largest contribution to the total uncertainty. There is observational evidence that the $f-$factor of $z>2$ galaxies is closer to $0.7$ than the locally measured value of $0.44$ \citep[][]{KASHINO17,FAISST19b,RODRIGUEZMUNOZ21}. In the case of $f=0.7$, the dust attenuation factors would decrease by a factor of $1.5$ ($0.18\,{\rm dex}$ change in luminosity) on average for our sample.

Finally, the right panel of Figure~\ref{fig:sfr_oii} relates the observed (solid symbols) and intrinsic (dust-corrected, empty symbols) \oii~luminosities to the total star formation (as described in Section~\ref{sec:discussionTotalSFR}). The far-infrared detected galaxies (dust correction from the IRX ratio) are shown as triangles, the remaining galaxies are shown as circles. \thegalaxy~is denoted with a star and the median of the dust-corrected \oii~luminosity and total SFR is marked as a purple cross.
The measurement uncertainties in \oii~luminosities are large compared to the measurement uncertainties in total SFR (on the order of $0.1\,{\rm dex}$). However, we expect systematic errors in the total SFRs to be a factor of $2-3$ larger given the uncertainties in the relation between \cii~and SFR.

Along with our data, we show different parameterisations of the \oii$-$SFR relation from the literature.
We include the original relation by \citet[][]{KENNICUTT98}\footnote{Converted to a \citet[][]{CHABRIER03} IMF.},
\begin{equation}\label{eq:oii_K98}
    {\rm SFR_{\oii}^{\rm K98}\,(M_{\odot}\,yr^{-1})} = (0.79 \pm 0.23) \times 10^{-41} L_{\rm [OII]} ({\rm erg\,s^{-1}}),
\end{equation}
as well as an updated version provided by equation 4 in \citet[][]{KEWLEY04}\footnote{Converted to a \citet[][]{CHABRIER03} IMF.}
\begin{equation}\label{eq:oii_K04}
    {\rm SFR_{\oii}^{\rm K04}\,(M_{\odot}\,yr^{-1})} = (3.72 \pm 0.93) \times 10^{-42} L_{\rm [OII]} ({\rm erg\,s^{-1}}).
\end{equation}
As pointed out by these authors, this relation may depend significantly on the gas-phase metallicity of the galaxies. We therefore also show their theoretical relations for solar, half-solar, and one-fifth solar metallicity obtained by equation 10 in \citet[][]{KEWLEY04}\footnote{Converted to a \citet[][]{CHABRIER03} IMF.}
\begin{equation}\label{eq:oii_K04_met}
    {\rm SFR_{\oii}^{\rm K04}}(Z)\,{\rm (M_{\odot}\,yr^{-1})} = \frac{4.46 \times 10^{-42} L_{\rm [OII]} ({\rm erg\,s^{-1}})}{(-1.75\pm0.25)[\log({\rm O/H})+12] + (16.73\pm2.23)}.
\end{equation}

We find that the original \citet[][]{KENNICUTT98} relation would significantly overestimate \oii-derived SFRs by factors of $3-5$ if applied to \textit{dust-corrected} \oii~luminosities. Note that the relation in Equation~\ref{eq:oii_K98} required dust corrections of \oii~at the \halpha~rest-frame wavelength \citep[see description in][]{KENNICUTT98}. However, at the dust attenuation values of our galaxies, this would reduce the dust correction factors by less than $40\%$ (and similarly the overestimation), hence cannot account for the discrepancy.
A much better estimate of the total SFR is provided by the updated and metal-dependent \oii$-$SFR relations by \citet[][]{KEWLEY04} (detailed numbers for the case of \thegalaxy~are listed in Table~\ref{tab:sfrs}). The large uncertainties, mainly due to the unknown differential dust attenuation factors but also the uncertain total SFR, do not allow us to distinguish relations for different metallicities at significance. As shown by the purple cross denoting the median value of the dust corrected \oii~luminosities and total SFRs, sub-solar metallicities are the most likely choice but solar metallicites cannot be excluded given the combination of measurement and systematic uncertainties.
Also, note that an increase of the $f-$factor from local $0.44$ to $\sim0.7$ would reduce the dust correction for the \oii~luminosity by $<0.2\,{\rm dex}$ (see above), which would keep our observations consistent with the \citep[][]{KEWLEY04} relation at sub-solar metallicity.

\begin{figure*}
    \centering
    \includegraphics[width=0.9\textwidth]{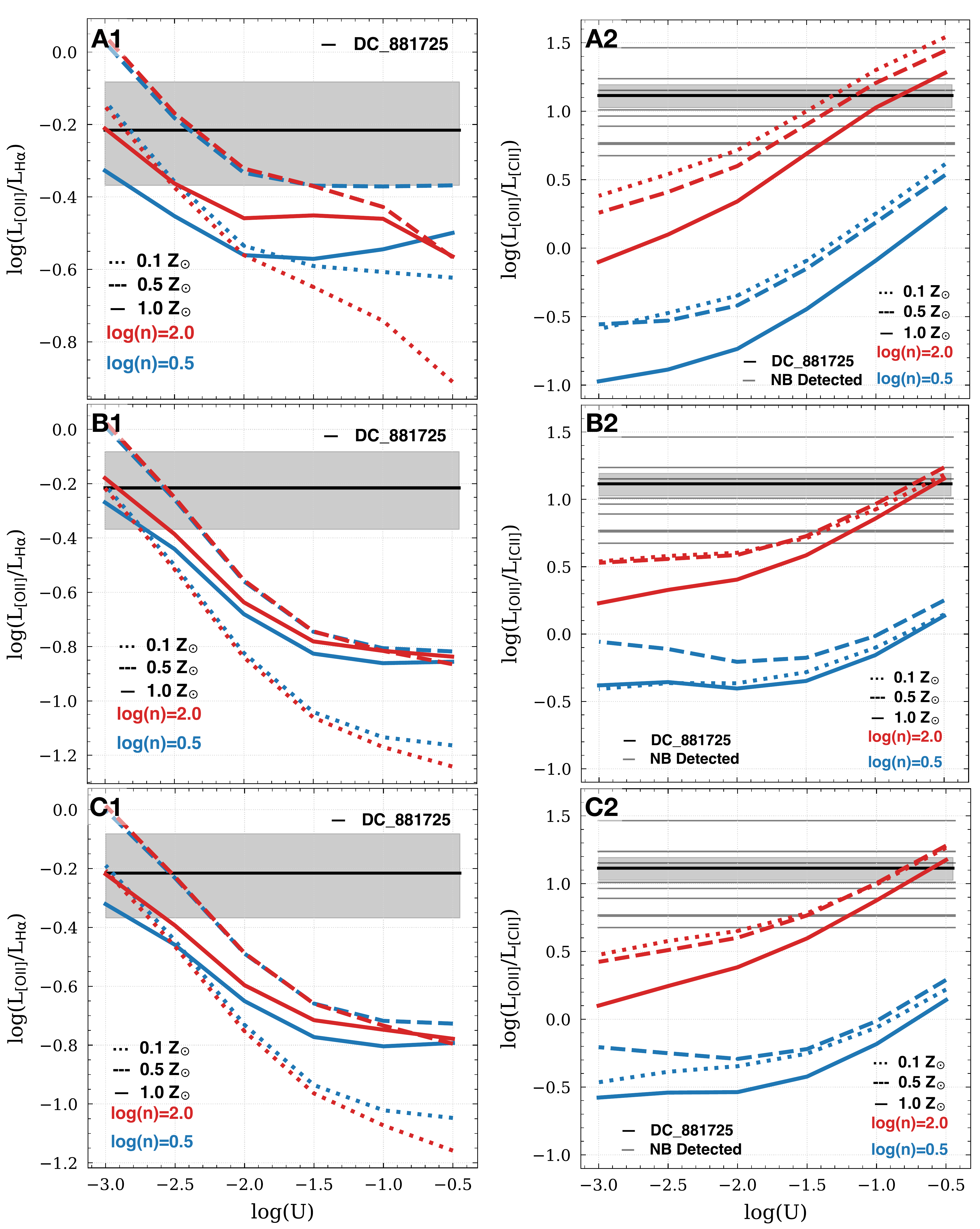}
    \caption{Predictions of the emission line ratio $\oiiha$ (left) and $\oiicii$ (right) from \textsc{Cloudy} simulations as a function of ionisation parameter ($U$) for $0.1\,{\rm Z_{\odot}}$ (dotted lines), $0.5\,{\rm Z_{\odot}}$ (dashed lines), and $1\,{\rm Z_{\odot}}$ (solid lines) metallicities as well as hydrogen densities of $\log(n / [{\rm cm^3}])=0.5$ (blue) and $\log(n / [{\rm cm^3}])=2$ (red).
    The different panels show different assumptions for the background stellar population (see text): burst of star formation (panels $A$), constant star formation ($B$), and best-fit SED for \thegalaxy~($C$).
    Our observed (dust-corrected) line ratios are shown as horizontal lines. In the case of $\oiicii$ (right panels), we show \thegalaxy~in black (with the $1\sigma$ uncertainty shown as grey swath) and the $9$ narrow-band detected galaxies as grey lines. For $\oiiha$ (left panels), we only show the median line ratio marginalised over all the uncertainties of the individual galaxies ($1\sigma$ uncertainties shown by the grey swath). Note that the $\oiiha$ ratio is largely insensitive to the hydrogen density, hence provides a good constraint on $U$. 
    \label{fig:CLOUDY}}
\end{figure*}

\subsection{Constraints on ISM Properties from \textsc{Cloudy} Analysis}\label{sec:discussioncloudy}

In this section, we compare the $\oiicii$~and $\oiiha$~luminosity ratios from measurements presented in Sections~\ref{sec:oii_measurements} and~\ref{sec:ha_measurements} to theoretical values obtained by \textsc{Cloudy} \citep[C17.02;][]{FERLAND17} to study key parameters of the ISM of our galaxies. 

For the \textsc{Cloudy} simulation, we assume gas with plain-parallel geometry including Orion-type grains as well as Polycyclic Aromatic Hydrocarbons (PAHs).
We ran models for gas clouds with electron densities of $\log(n_{\rm H}/{\rm cm^{-3}})\,=\,[0.5,2]$, gas-phase metallicities of $Z=[0.1,0.5,1.0]\,{\rm Z_{\odot}}$, and ionisation parameters between $-3.0 < \log(U) < -0.5$ in steps of $0.5\,{\rm dex}$.
The stopping criteria of the simulation is crucial, especially for computing the \cii~emission, which can originate from low-density warm ISM. We use a stopping point at $A_{\rm v}=100\,{\rm mag}$ and we found that lower values such as $A_{\rm v}=10\,{\rm mag}$ do not have a significant impact on the our results.
We also investigated the dependence of our results on different star formation histories for the underlying stellar population. For this, we assumed three different simple models: \textit{(A)} a starburst with an age of $100\,{\rm Myr}$; \textit{(B)} a constant star formation history; \textit{(C)} the best-fit SED of \thegalaxy. 
The first two are produced by the \textsc{BPASS} v2.0 models \citep[][]{ELDRIDGE17,STANWAY18}, assume $10\%$ solar stellar metallicity, and include binary-star evolution. The latter is derived from broad-band photometry using \textsc{CIGALE} \citep[][]{BURGARELLA05,NOLL09,BOQUIEN19}.
The results are shown in Figure~\ref{fig:CLOUDY}. The panel labels indicate the assumption of the underlying stellar population (\textit{A}, \textit{B}, or \textit{C}, respectively). The left and right columns show the result for $\oiiha$~and $\oiicii$, respectively.
In the following, we compare the models to the observations in more detail.

\subsubsection{The $\oiiha$~luminosity ratio}\label{sec:oiiha}

Let us first have a look at the $\oiiha$~luminosity ratio. As seen on the left panels in Figure~\ref{fig:CLOUDY}, this ratio shows a strong negative correlation with the ionisation parameter (radiation pressure) $U$, however is largely independent of the electron density and the changes with metallicity are small compared to the accuracy of our observations. Thus the $\oiiha$~luminosity ratio allows us to put constraints on the ionisation parameter $U$. The trends are slightly flatter for a constant star formation history or using the best-fit SED. However, overall the impact of the three different star formation histories is negligible at the current accuracy of our measurements.

The observed median dust-corrected $\oiiha$ ratio ($-0.22_{-0.15}^{+0.13}$) and population scatter is indicated as black line with grey area, respectively. Note that we are only using $7$ out of the $10$ galaxies whose \textit{Spitzer} photometry is not contaminated by nearby sources. \textit{Spitzer}-derived \halpha~luminosities are mainly uncertain because of the unknown dust correction (specifically the unknown $f-$factor). However, the \oii~to~\halpha~luminosity ratio is expected to be rather stable as the \oii~and \halpha~lines are affected by dust in a similar way (and have likely similar $f-$factors). For example, a change in $f$ from $0.44$ to $0.7$ affects the luminosity ratio by less than $0.1\,{\rm dex}$, which is smaller than sample scatter and measurement uncertainties. Different reddening curves contribute even less to the uncertainties.

Comparing our observations to the \textsc{Cloudy} models, we find that low ionisation parameters ($\log(U) < -2$ or $\log(q / {\rm cm\,s^{-1}}) < 8.5$)\footnote{Note that $U = q/c$, where $c=3\times10^{10}\,{\rm cm\,s^{-1}}$ is the speed of light.} are favored independent of the assumed electron density and metallicity.
Such ionisation parameter values are very consistent with typical measurements at $z=2-3$ (via the \oiii/\oii~line ratio) finding $\log(U) \sim -2.5$ or $\log(q / {\rm cm\,s^{-1}}) \sim 8$ \citep[][]{NAKAJIMA14,SANDERS16,SANDERS20}.
On the other hand, studies at $z>6$ suggest $\log(U) \sim -1.7$, which would indicate an increase in ionisation parameter in the Epoch of Reionisation \citep[][]{STARK14,STARK15,HUTCHISON19,HARIKANE20}.
Note that observations of local \hii~regions suggest an upper cut-off due to radiation confinement of $\log(U)~\sim~-1$ \citep[e.g.,][]{YEH12}.

\subsubsection{The $\oiicii$~luminosity ratio}\label{sec:oiicii}

Next, we focus on the $\oiicii$~luminosity ratio. The models from \textsc{Cloudy} are shown in the right panels of Figure~\ref{fig:CLOUDY}. Increasing ionisation parameter as well as increasing electron density result in a higher luminosity ratio. On the other hand, more metal enriched environments are expected to have lower $\oiicii$ ratios. The trends are not significantly affected by the different assumptions on the underlying stellar population.

The black and grey horizontal lines show the measured ratios for \thegalaxy~and the $9$ narrow-band detected galaxies, respectively. The \oii~luminosities have been corrected for dust attenuation using the methods described in Section~\ref{sec:discussionOIISFR}.
We find a log-ratio of $1.11_{-0.08}^{+0.08}$ for \thegalaxy~ and a median ratio of $0.98^{+0.21}_{-0.22}$ for all $10$ galaxies.
Note that the true stellar reddening curve plays a sub-dominant role in this rather quantitative analysis.

The models show that the intrinsic $\oiicii$ luminosity ratio is significantly degenerate with metallicity, ionisation parameter, and electron density.
From Section~\ref{sec:oiiha} we expect an ionisation parameter $\log(U) < -2$.
Metallicity seems to have the least impact on the luminosity ratio assuming a reasonable range between $10\%$ solar and solar metallicity. An analysis of the rest-frame UV absorption line strengths (see Section~\ref{sec:metallicity}) suggests that our galaxies have $\sim50\%$ of the solar gas-phase metal enrichment on average. Fixing the metallicity to half-solar is therefore a reasonable assumption.
Within these assumptions our measurements would argue for electron densities of $\log(n/{\rm [cm^{-3}]}) \sim 2.5-3$,

Typical electron densities are measured to be around $100-200\,{\rm cm^{-3}}$ in galaxies with similar rates of star formation at $z\sim2$. This is about a factor of $5-10$ higher than observed on average in local galaxies \citep[][]{STEIDEL14,MASTERS14,SANDERS16,DAVIES21}. However, as pointed out by \citet[][]{KAASINEN17} \citep[see also][]{SHIMAKAWA15,DAVIES21}, the electron density is positively correlated with (specific) SFR, hence the higher $\log(n)$ found in high-$z$ galaxies are likely due to a higher normalization of the main-sequence (i.e., higher average SFR). In line with this, it was found that $\log(n)$ does not significantly evolve with redshift for SFR-matched samples.
Taking the measurements in \citet[][]{DAVIES21} at face value suggests at least $\log(n/{\rm cm^{-3}}) \sim 2.5$ ($300\,{\rm cm^{-3}}$) for the average specific SFR of our sample ($\sim0.9\,{\rm Gyr^{-1}}$) according to $z\sim2$ galaxies.
Our measurements ($\log(n/{\rm [cm^{-3}]}) \sim 2.5-3$) are generally consistent with these expectations \textit{given their high SFRs}, although at the high end. Lower electron densities (e.g., $\sim100-200\,{\rm cm^{-3}}$, typical at $z\sim2$) would indicate ionisation parameters of $\log(U) \gtrsim -1.7$, which is clearly inconsistent with our measured $\oiiha$~luminosity ratio.

Combining the results of this section with Section~\ref{sec:oiiha}, we conclude that our $z\sim4.5$ galaxies have ionisation parameters $\log(U) <  -2$ and electron densities $\log(n/{\rm [cm^{-3}]}) \sim 2.5-3.0$. While the former is consistent with measurements in galaxies at $z=2-3$, the latter is consistent with or slightly higher than the expected electron densities given the SFRs of our galaxies and the relation between $\log(n)$ and star formation measured in lower redshift galaxies.
Significantly higher ionisation parameters are suggested for galaxies in the Epoch of Reionisation at $z>6$, which would argue for a fast evolution of the ISM properties at times earlier than $z\sim4.5$.
Furthermore, we find that these typical main-sequence galaxies are much less extreme than sub-millimetre galaxies at similar redshifts showing electron densities of $\log(n/{\rm [cm^{-3}]}) = 4$ or more \citep[e.g.,][]{BOTHWELL17,DEBREUCK19}.

We note that nine of the galaxies lie in a greater structure associated with a proto-cluster at $z\sim4.57$ \citep[][]{Lemaux2018}. It is possible that the proto-cluster environment has altered their properties in some way that could be relevant for this study. For example, \citet[][]{DARVISH15} argues for a lower electron density in galaxies residing in dense environments ($\Sigma/\Sigma_{\rm mean} > 4$) based on a comprehensive study of an over dense region at $z\sim0.5$. The translation of this result to (less) over-dense regions at $z\sim5$ and hence the impact of the environment on the properties of our galaxies remains unclear. As far as we can tell, the nine galaxies have statistically indistinguishable properties with respect to the other \textit{ALPINE} (hence field) galaxies using the metrics that we are able to compare (e.g., Figure~\ref{fig:mainsequence}). As such, it is at least likely that any effect of environment on the ISM is subtle at best.

\subsubsection{Caveats}
Several other parameters could change the modeled line ratios. Here, we investigate the impact of the stellar metallicity of the stellar population producing the incident spectrum as well as the geometrical covering factor.

First, we ran the simulation with a covering factor of $0.1$ instead of $1$ and found that it does not have a significant impact ($<0.05\,{\rm dex}$) on either line ratio.
Second, we implemented a solar stellar metallicity (instead of $10\%$ solar) for the incident spectrum produced by the \textsc{BPASS} models. We find that this increases both $\oiiha$ and $\oiicii$ ratios by $0.15\,{\rm dex}$. This seems to be because a factor $8$ decrease in \oii, which is over-balanced by a factor $11$ and $12$ decrease in \halpha~and \cii, respectively.

An increase in the $\oiicii$ luminosity ratio would slightly reduce the likelihood of high electron densities for a given ionisation parameter. And increased $\oiiha$ ratio would further allow slightly higher ionisation parameters (especially if assuming a gas-phase metallicity of $50\%$ solar). All in all, a higher stellar metallicity would argue for $\sim 0.2-0.3\,{\rm dex}$ higher ionisation parameters and roughly the same amount lower electron densities, making our galaxies consistent with their counterparts at $z\sim2-3$ with similar star formation properties.

Finally, our simple model assumes that the galaxy is a single \hii~region. More complicated ``multi-sector'' models could be investigated, however, we do not think that given the uncertainties in our (photometric) measurements of \oii~and \halpha~this would improve the robustness of our results.

\section{Conclusions}\label{sec:end}

We have assembled the best-studied sample of main-sequence galaxies at $z\sim4.5$ to date, to investigate key parameters of the ISM as well as the relation between \oii~and total SFR of galaxies in the early Universe right after the Epoch of Reionisation.

The $10$ main-sequence galaxies at $z\sim4.5$ have measurements of rest-frame UV absorption lines, optical \oii~(from spectroscopy and narrow-band imaging) and \halpha~emission (from \textit{Spitzer}), as well as far-infrared observations of \cii~and dust continuum from ALMA.
We use the total SFRs derived from \cii~emission as well as far-infrared continuum to calibrate the relation between SFR and \oii~emission for the first time at these redshifts (Section~\ref{sec:discussionSFR}). Furthermore, we constrain key parameters of their ISM (metallicity, electron density, ionisation parameter) via the UV absorption lines and \oii/\cii~and \oii/\halpha~luminosity ratios, which we compare to \textsc{Cloudy} simulations (Sections~\ref{sec:metallicity} and~\ref{sec:discussioncloudy}).
Our findings can be summarised as follows:

\begin{itemize}
    \item The relation between dust-corrected \oii~luminosities and total SFR is best described by sub-solar metallicity models from \citet[][]{KEWLEY04}. For the median of our sample, we find a total SFR of $53.57^{+23.48}_{-31.30}\,{\rm M_{\odot}\,yr^{-1}}$ and $\log(L_{\rm [OII]} / {\rm L_{\odot}}) = 9.88^{+0.29}_{-0.51}$. The original \citet[][]{KENNICUTT98} \oii$-$SFR relation would overestimate the SFR by a factor of $\sim3$ (Section~\ref{sec:discussionOIISFR}).
    
    \item By comparing the $\oiicii$~and $\oiiha$~luminosity ratio to \textsc{Cloudy} models, we find that our galaxies have ionisation parameters $\log(U) < -2$ and electron densities of $\log(n/[{\rm cm^3}]) \sim 2.5-3$. The former is consistent with $z\sim2-3$ galaxies, the latter may be slightly higher than expected based on our galaxies' specific SFR. However, these results depend on the input parameters for the \textsc{Cloudy} simulation. Specifically, increasing the stellar metallicity of the incident spectrum from $10\%$ solar to solar would make our observations more consistent with electron densities measured in $z=2-3$ galaxies.
    
\end{itemize}

All in all, we find that the ISM properties of this representative sample of $z\sim4.5$ galaxies are similar to their descendants at $z\sim2-3$ when matched by their specific SFRs. There are indications that the ionisation parameter of more primordial galaxies in the Epoch of Reionisation at $z>6$ is increased \citep[see, e.g.,][]{HARIKANE20} compared to our findings. This suggests a rapid evolution of the ISM in the few $100\,{\rm Myrs}$ spanning these epochs.
The relation between \oii~emission and SFR is consistent if assuming a gas-phase metallicity of $\sim50\%$ solar, which is in line with the estimates from rest-frame UV absorption spectroscopy. This result is an important step towards using optical emission lines as total SFR indicators in the era of JWST.

In this work, we demonstrated the necessity the combination of data from many facilities covering the rest-frame UV to far-infrared to decipher the physical properties of high-redshift galaxies.
Although this is the largest and best-studied sample of typical $z\sim4.5$ galaxies to-date, larger samples with similar multi-wavelength observations are crucial for a better statistical study of the dependence of the reported parameters on other galaxy properties. Larger surveys with current facilities (e.g., \textit{Keck}) and later JWST will provide these samples.

\begin{table*}
\caption{Summary of measured properties for \thegalaxy~from broad-band photometry, ALMA observations, and new \textit{Keck}/MOSFIRE spectroscopy.} \label{tab:keckproperties}
\begin{tabular}{ccccc ccc cc cc}
  \hline\hline
  \multicolumn{5}{c}{\textbf{Broad-band}} & \multicolumn{3}{c}{\textbf{ALMA}} &
  \multicolumn{2}{c}{\textbf{\textit{Keck}/DEIMOS}} &
  \multicolumn{2}{c}{\textbf{\textit{Keck}/MOSFIRE}} \\[-0.1cm]
  \multicolumn{5}{c}{---------------------------------------------------------------------------} & \multicolumn{3}{c}{--------------------------------------------} &
  \multicolumn{2}{c}{---------------------------} &
  \multicolumn{2}{c}{---------------------------}\\[0.0cm]
  $\log({\rm M_*})$ & $\rm SFR_{\rm SED}$ & $\rm E(B-V)_{SED}$ & $\rm \log(L_{UV})$ & $\beta$ &$z_{\rm [CII]}$ & $\log(\rm L_{IR})$ & $\log(\rm L_{[CII]})$ & $z_{\rm Ly\alpha}$ & EW(Ly$\rm \alpha$) & $z_{\rm [OII]}$ & $\log(\rm L_{[OII]})$  \\[0.1cm]
   $\left[\rm{M_{\odot}}\right]$ & $\left[\rm{M_{\odot}\,yr^{-1}}\right]$ & $\left[\rm{mag}\right]$ & $\left[\rm{L_{\odot}}\right]$ & & & $\left[\rm{L_{\odot}}\right]$ & $\left[\rm{L_{\odot}}\right]$ &  & [\AA] &  & $\left[\rm{L_{\odot}}\right]$  \\ \hline
   9.96$^{+0.16}_{-0.11}$ & 88.0$^{+61.1}_{-43.3}$ & 0.25$^{+0.05}_{-0.05}$ & 10.97$^{+0.06}_{-0.06}$ & -1.20$^{+0.42}_{-0.18}$ & 4.5777 & 11.67$^{+0.10}_{-0.13}$ & 8.84$^{+0.04}_{-0.04}$ & 4.5854 & 57.4$^{+18.0}_{-18.0}$ & 4.5793 & 9.22$^{+0.05}_{-0.06}$\\
        \hline
 \end{tabular}\\
\end{table*}

\begin{table*}
\caption{Summary of properties derived from the ancillary data for the $9$ galaxies with \oii~detection from narrow-band imaging.} \label{tab:narrowbandproperties}
\begin{tabular}{lccccc ccc}
  \hline\hline
  ID & z$_{\rm [CII]}$ & log(L$_{\rm {OII}}$) & log(L$_{\rm{[CII]}}$)& log(L$_{\rm FIR}$)& log(M)& SFR$_{\rm [CII]}$ & SFR$_{\rm FIR}$ & SFR$_{\rm UV}$\\
   &  & $\left[\rm{L_{\odot}}\right]$ & $\left[\rm{L_{\odot}}\right]$ & $\left[\rm{L_{\odot}}\right]$ & $\left[\rm{M_{\odot}}\right]$ & $\left[\rm{M_{\odot}\,yr^{-1}}\right]$ & $\left[\rm{M_{\odot}\,yr^{-1}}\right]$ & $\left[\rm{M_{\odot}\,yr^{-1}}\right]$  \\ \hline
   DC\_665626 & 4.5830 & 9.36$^{+0.08}_{-0.09}$ & 8.21$^{+0.10}_{-0.13}$&--&9.21$^{+0.16}_{-0.18}$&23.15$^{+1.22}_{-1.30}$&--&5.77$^{+1.24}_{-1.23}$\\[0.1cm]
   DC\_680104 & 4.5320 & 9.51$^{+0.10}_{-0.13}$ & 8.74$^{+0.12}_{-0.16}$&--&9.23$^{+0.18}_{-0.12}$&66.30$^{+1.25}_{-1.36}$&--&14.55$^{+1.12}_{-1.10}$\\[0.1cm]
   VC\_5100969402 & 4.5869 & 9.34$^{+0.08}_{-0.10}$ & 8.72$^{+0.04}_{-0.05}$&11.65$^{+0.11}_{-0.16}$&10.00$^{+0.14}_{-0.12}$&63.42$^{+1.09}_{-1.10}$&43.23$^{+13.06}_{-13.06}$&12.50$^{+1.14}_{-1.13}$\\[0.1cm]
   VC\_5100994794 & 4.5783 & 9.14$^{+0.11}_{-0.15}$ & 8.75$^{+0.04}_{-0.04}$&11.20$^{+0.12}_{-0.16}$&9.73$^{+0.13}_{-0.15}$&66.93$^{+1.08}_{-1.09}$&15.50$^{+4.71}_{-4.71}$&10.69$^{+1.19}_{-1.23}$\\[0.1cm]
   VC\_5101209780 & 4.5700 & 8.99$^{+0.30}_{-0.38}$ & 8.86$^{+0.08}_{-0.10}$&11.62$^{+0.13}_{-0.19}$&10.05$^{+0.12}_{-0.12}$&84.45$^{+1.18}_{-1.23}$&40.65$^{+14.64}_{-14.64}$&22.07$^{+1.09}_{-1.10}$\\[0.1cm]
   VC\_5101210235 & 4.5733 & 9.06$^{+0.16}_{-0.30}$ & 8.35$^{+0.11}_{-0.14}$&--&9.78$^{+0.15}_{-0.12}$&30.71$^{+1.24}_{-1.33}$&--&24.08$^{+1.08}_{-1.07}$\\[0.1cm]
   VC\_5101218326 & 4.5678 & 9.32$^{+0.08}_{-0.10}$ & 9.26$^{+0.02}_{-0.02}$&11.79$^{+0.07}_{-0.08}$&11.01$^{+0.05}_{-0.07}$&184.45$^{+1.04}_{-1.04}$&60.38$^{+10.30}_{-10.30}$&27.94$^{+1.06}_{-1.07}$\\[0.1cm]
   VC\_5101244930 & 4.5769 & 9.09$^{+0.30}_{-0.19}$ & 8.70$^{+0.08}_{-0.10}$&--&9.67$^{+0.13}_{-0.16}$&61.73$^{+1.18}_{-1.23}$&--&17.45$^{+1.12}_{-1.12}$\\[0.1cm]
   VC\_5110377875 & 4.5441 & 9.21$^{+0.18}_{-0.32}$ & 9.23$^{+0.03}_{-0.03}$&--&10.17$^{+0.21}_{-0.12}$&174.87$^{+1.05}_{-1.06}$&--&24.67$^{+1.09}_{-1.08}$\\
        \hline
 \end{tabular}\\
  \begin{tablenotes}
   \item \textbf{Notes:} The UV SFRs have not been corrected for dust attenuation. \cii~SFRs derived using the \citet[][]{ALPINE_SCHAERER20} relation (see Section~\ref{sec:discussionTotalSFR}). Far-infrared SFRs derived using the \citet[][]{KENNICUTT98} relation.
  \end{tablenotes}
\end{table*}


\begin{table}
\caption{Summary of SFRs measured from \oii~ and ancillary \textit{ALPINE} data available for \thegalaxy. Note that SFRs from UV and optical indicators have not been corrected for dust attenuation.} \label{tab:sfrs}
\begin{tabular}{l cccc}
  \hline\hline
  & \multicolumn{4}{c}{\textbf{SFR $[\rm M_{\astrosun}\,yr^{-1}]$ }}  \\[-0.1cm]
  & \multicolumn{4}{c}{-----------------------------------------------------------}\\[0.0cm]
  Reference & \oii & \cii & IR & UV \\ \hline
   \textbf{K98}      & 50.0$^{+6.7}_{-6.1}$ &     & 46.0$^{+11.9}_{-11.9}$ & 13.3$^{+2.0}_{-1.8}$\\[0.1cm]
   \textbf{K04}$^\dagger$      &   &     &   &  \\[0.1cm]
   - no met. dep.      & 23.6$^{+3.2}_{-2.7}$ &     &   &  \\[0.1cm]
   - solar      & 18.6$^{+2.5}_{-2.3}$ &     &   &  \\[0.1cm]
   - half-solar      & 13.8$^{+1.9}_{-1.7}$ &     &   &  \\[0.1cm]
   - $1/5^{\rm th}$-solar      & 10.3$^{+1.4}_{-1.5}$ &     &   &  \\[0.1cm]
    \textbf{L14}        &      &                                & & \\[0.1cm]
   - metal-poor dwarfs                &      & 22.0$^{+1.6}_{-1.6}$     & & \\[0.1cm]
   - starburst                &      & 59.9$^{+5.5}_{-5.5}$      & & \\[0.1cm]
   - AGN                      &      & 73.0$^{+6.0}_{-6.1}$      & & \\[0.1cm]
   - high-$z$                 &      & 80.9$^{+8.9}_{-8.7}$        & & \\[0.1cm]
   - ULIRGs                   &      & 360.7$^{+4.3}_{-4.3}$       &  &\\[0.1cm]
   \textbf{S20}    &      & 64.1$^{+6.2}_{-5.7}$      &  &\\
   \hline
  dust correction factor$^\ddagger$  &   5.2 (4.3)   &   $-$     & $-$ & 3.4 (3.5)\\[0.1cm]
        \hline
 \end{tabular}\\
  \begin{tablenotes}
  \item $^\dagger$ The first value is derived using their equation 4. The other values are derived for different gas-phase metallicities and their equation 10.
  \item $^\ddagger$ Dust correction factor for \oii~and UV continuum emission assuming a reddening curve from \citet[][]{CALZETTI00} and \citet[][]{REDDY15} (in parenthesis), respectively (see text for more details). The \oii~correction factor assumes a differential dust attenuation $f=0.44$. For $f=0.7$, decrease the value by a factor of $1.8$. 
   \item \textbf{References:} K98 $-$ \citet[][]{KENNICUTT98};
   K04 $-$ \citet[][]{KEWLEY04};
   L14 $-$ \citet[][]{DELOOZE14};
   S20 $-$ \citet[][]{ALPINE_SCHAERER20}.
  \end{tablenotes}
\end{table}

\section*{Acknowledgements}

We thank Yuichi Harikane for the helpful inputs on running \textsc{Cloudy} and the anonymous referee for the comments which improved this manuscript. We also thank Behnam Darvish and Nick Scoville for letting us add \thegalaxy~to their \textit{MOSFIRE} mask. G.C.J. acknowledges ERC Advanced Grant 695671 ``QUENCH'' and support by the Science and Technology Facilities Council (STFC).
This work made use of v2.2.1 of the Binary Population and Spectral Synthesis (BPASS) models as described in \citet[][]{ELDRIDGE17} and \citet[][]{STANWAY18}.
This work was supported by the Programme National Cosmology et Galaxies (PNCG) of CNRS/INSU with INP and IN2P3, co-funded by CEA and CNES.
This paper uses data obtained with
the ALMA Observatory, under Large Program \textit{2017.1.00428.L}. ALMA is a partnership of ESO (representing its member states), NSF (USA) and NINS (Japan), together with NRC (Canada), MOST and ASIAA (Taiwan), and KASI (Republic of Korea), in cooperation with the Republic of Chile. The Joint ALMA Observatory is operated by ESO, AUI/NRAO and NAOJ.
This work is based on observations and archival data made with the \textit{Spitzer Space Telescope}, which is operated by the Jet Propulsion Laboratory, California Institute of Technology, under a contract with NASA.
Some of the material presented in this paper is based upon work supported by the National Science Foundation under Grant No. 1908422. 
MT acknowledges the support from grant PRIN MIUR 2017 20173ML3WW 001.
Furthermore, this work is based on data from the W. M. Keck Observatory and the Subaru Telescope. The authors wish to recognize and acknowledge the very significant cultural role and reverence that the summit of Mauna Kea has always had within the indigenous Hawaiian community. We are most fortunate to have the opportunity to conduct observations from
this mountain.
Based on data products from observations made with ESO Telescopes at the La Silla Paranal Observatory under ESO programme ID 179.A-2005 and on data products produced by CALET and the Cambridge Astronomy Survey Unit on behalf of the UltraVISTA consortium.
Finally, we would also like to recognize the contributions from all of the members of the COSMOS Team who helped in obtaining and reducing the large amount of multi-wavelength data that are now publicly available through IRSA at \url{http://irsa.ipac.caltech.edu/Missions/cosmos.html}.

\section*{Data Availability}
The data underlying this article will be shared on reasonable request to the corresponding author.




\bibliographystyle{mnras}
\bibliography{bibli} 

\begin{thebibliography}{}
\makeatletter
\relax
\def\mn@urlcharsother{\let\do\@makeother \do\$\do\&\do\#\do\^\do\_\do\%\do\~}
\def\mn@doi{\begingroup\mn@urlcharsother \@ifnextchar [ {\mn@doi@}
  {\mn@doi@[]}}
\def\mn@doi@[#1]#2{\def\@tempa{#1}\ifx\@tempa\@empty \href
  {http://dx.doi.org/#2} {doi:#2}\else \href {http://dx.doi.org/#2} {#1}\fi
  \endgroup}
\def\mn@eprint#1#2{\mn@eprint@#1:#2::\@nil}
\def\mn@eprint@arXiv#1{\href {http://arxiv.org/abs/#1} {{\tt arXiv:#1}}}
\def\mn@eprint@dblp#1{\href {http://dblp.uni-trier.de/rec/bibtex/#1.xml}
  {dblp:#1}}
\def\mn@eprint@#1:#2:#3:#4\@nil{\def\@tempa {#1}\def\@tempb {#2}\def\@tempc
  {#3}\ifx \@tempc \@empty \let \@tempc \@tempb \let \@tempb \@tempa \fi \ifx
  \@tempb \@empty \def\@tempb {arXiv}\fi \@ifundefined
  {mn@eprint@\@tempb}{\@tempb:\@tempc}{\expandafter \expandafter \csname
  mn@eprint@\@tempb\endcsname \expandafter{\@tempc}}}

\bibitem[\protect\citeauthoryear{{Ando}, {Ohta}, {Iwata}, {Akiyama}, {Aoki}  \&
  {Tamura}}{{Ando} et~al.}{2007}]{ANDO07}
{Ando} M.,  {Ohta} K.,  {Iwata} I.,  {Akiyama} M.,  {Aoki} K.,   {Tamura} N.,
  2007, \mn@doi [\pasj] {10.1093/pasj/59.4.717}, \href
  {http://adsabs.harvard.edu/abs/2007PASJ...59..717A} {59, 717}

\bibitem[\protect\citeauthoryear{{Arnouts}, {Cristiani}, {Moscardini},
  {Matarrese}, {Lucchin}, {Fontana}  \& {Giallongo}}{{Arnouts}
  et~al.}{1999}]{ARNOUTS99}
{Arnouts} S.,  {Cristiani} S.,  {Moscardini} L.,  {Matarrese} S.,  {Lucchin}
  F.,  {Fontana} A.,   {Giallongo} E.,  1999, \mn@doi [\mnras]
  {10.1046/j.1365-8711.1999.02978.x}, \href
  {http://adsabs.harvard.edu/abs/1999MNRAS.310..540A} {310, 540}

\bibitem[\protect\citeauthoryear{{Asplund}, {Grevesse}, {Sauval}  \&
  {Scott}}{{Asplund} et~al.}{2009}]{ASPLUND09}
{Asplund} M.,  {Grevesse} N.,  {Sauval} A.~J.,   {Scott} P.,  2009, \mn@doi
  [\araa] {10.1146/annurev.astro.46.060407.145222}, \href
  {https://ui.adsabs.harvard.edu/abs/2009ARA&A..47..481A} {47, 481}

\bibitem[\protect\citeauthoryear{{Bertin}}{{Bertin}}{2006}]{Bertin2006}
{Bertin} E.,  2006, in {Gabriel} C.,  {Arviset} C.,  {Ponz} D.,   {Enrique} S.,
   eds,  Astronomical Society of the Pacific Conference Series Vol. 351,
  Astronomical Data Analysis Software and Systems XV. p.~112

\bibitem[\protect\citeauthoryear{{Bertin}}{{Bertin}}{2011}]{Bertin2011}
{Bertin} E.,  2011, in {Evans} I.~N.,  {Accomazzi} A.,  {Mink} D.~J.,   {Rots}
  A.~H.,  eds,  Astronomical Society of the Pacific Conference Series Vol. 442,
  Astronomical Data Analysis Software and Systems XX. p.~435

\bibitem[\protect\citeauthoryear{{Bertin} \& {Arnouts}}{{Bertin} \&
  {Arnouts}}{1996}]{Bertin1996}
{Bertin} E.,  {Arnouts} S.,  1996, \mn@doi [\aaps] {10.1051/aas:1996164}, \href
  {https://ui.adsabs.harvard.edu/abs/1996A&AS..117..393B} {117, 393}

\bibitem[\protect\citeauthoryear{{Bertin}, {Mellier}, {Radovich}, {Missonnier},
  {Didelon}  \& {Morin}}{{Bertin} et~al.}{2002}]{Bertin2002}
{Bertin} E.,  {Mellier} Y.,  {Radovich} M.,  {Missonnier} G.,  {Didelon} P.,
  {Morin} B.,  2002, in {Bohlender} D.~A.,  {Durand} D.,   {Handley} T.~H.,
  eds,  Astronomical Society of the Pacific Conference Series Vol. 281,
  Astronomical Data Analysis Software and Systems XI. p.~228

\bibitem[\protect\citeauthoryear{{B{\'e}thermin} et~al.,}{{B{\'e}thermin}
  et~al.}{2020}]{ALPINE_BETHERMIN19}
{B{\'e}thermin} M.,  et~al., 2020, \mn@doi [\aap]
  {10.1051/0004-6361/202037649}, \href
  {https://ui.adsabs.harvard.edu/abs/2020A&A...643A...2B} {643, A2}

\bibitem[\protect\citeauthoryear{{Boquien}, {Burgarella}, {Roehlly}, {Buat},
  {Ciesla}, {Corre}, {Inoue}  \& {Salas}}{{Boquien} et~al.}{2019}]{BOQUIEN19}
{Boquien} M.,  {Burgarella} D.,  {Roehlly} Y.,  {Buat} V.,  {Ciesla} L.,
  {Corre} D.,  {Inoue} A.~K.,   {Salas} H.,  2019, \mn@doi [\aap]
  {10.1051/0004-6361/201834156}, \href
  {https://ui.adsabs.harvard.edu/abs/2019A&A...622A.103B} {622, A103}

\bibitem[\protect\citeauthoryear{{Bothwell} et~al.,}{{Bothwell}
  et~al.}{2017}]{BOTHWELL17}
{Bothwell} M.~S.,  et~al., 2017, \mn@doi [\mnras] {10.1093/mnras/stw3270},
  \href {https://ui.adsabs.harvard.edu/abs/2017MNRAS.466.2825B} {466, 2825}

\bibitem[\protect\citeauthoryear{{Bruzual} \& {Charlot}}{{Bruzual} \&
  {Charlot}}{2003}]{BRUZUALCHARLOT03}
{Bruzual} G.,  {Charlot} S.,  2003, \mn@doi [\mnras]
  {10.1046/j.1365-8711.2003.06897.x}, \href
  {http://adsabs.harvard.edu/abs/2003MNRAS.344.1000B} {344, 1000}

\bibitem[\protect\citeauthoryear{{Buat} et~al.,}{{Buat} et~al.}{2005}]{BUAT05}
{Buat} V.,  et~al., 2005, \mn@doi [\apjl] {10.1086/423241}, \href
  {http://adsabs.harvard.edu/abs/2005ApJ...619L..51B} {619, L51}

\bibitem[\protect\citeauthoryear{{Burgarella}, {Buat}  \&
  {Iglesias-P{\'a}ramo}}{{Burgarella} et~al.}{2005}]{BURGARELLA05}
{Burgarella} D.,  {Buat} V.,   {Iglesias-P{\'a}ramo} J.,  2005, \mn@doi
  [\mnras] {10.1111/j.1365-2966.2005.09131.x}, \href
  {http://adsabs.harvard.edu/abs/2005MNRAS.360.1413B} {360, 1413}

\bibitem[\protect\citeauthoryear{{Calzetti}, {Armus}, {Bohlin}, {Kinney},
  {Koornneef}  \& {Storchi-Bergmann}}{{Calzetti} et~al.}{2000}]{CALZETTI00}
{Calzetti} D.,  {Armus} L.,  {Bohlin} R.~C.,  {Kinney} A.~L.,  {Koornneef} J.,
   {Storchi-Bergmann} T.,  2000, \mn@doi [\apj] {10.1086/308692}, \href
  {http://adsabs.harvard.edu/abs/2000ApJ...533..682C} {533, 682}

\bibitem[\protect\citeauthoryear{{Chabrier}}{{Chabrier}}{2003}]{CHABRIER03}
{Chabrier} G.,  2003, \mn@doi [\pasp] {10.1086/376392}, \href
  {http://adsabs.harvard.edu/abs/2003PASP..115..763C} {115, 763}

\bibitem[\protect\citeauthoryear{{Croxall} et~al.,}{{Croxall}
  et~al.}{2017}]{CROXALL17}
{Croxall} K.~V.,  et~al., 2017, \mn@doi [\apj] {10.3847/1538-4357/aa8035},
  \href {https://ui.adsabs.harvard.edu/abs/2017ApJ...845...96C} {845, 96}

\bibitem[\protect\citeauthoryear{{Darvish}, {Mobasher}, {Sobral}, {Hemmati},
  {Nayyeri}  \& {Shivaei}}{{Darvish} et~al.}{2015}]{DARVISH15}
{Darvish} B.,  {Mobasher} B.,  {Sobral} D.,  {Hemmati} S.,  {Nayyeri} H.,
  {Shivaei} I.,  2015, \mn@doi [\apj] {10.1088/0004-637X/814/2/84}, \href
  {https://ui.adsabs.harvard.edu/abs/2015ApJ...814...84D} {814, 84}

\bibitem[\protect\citeauthoryear{{Davies} et~al.,}{{Davies}
  et~al.}{2021}]{DAVIES21}
{Davies} R.~L.,  et~al., 2021, \mn@doi [\apj] {10.3847/1538-4357/abd551}, \href
  {https://ui.adsabs.harvard.edu/abs/2021ApJ...909...78D} {909, 78}

\bibitem[\protect\citeauthoryear{{De Breuck} et~al.,}{{De Breuck}
  et~al.}{2019}]{DEBREUCK19}
{De Breuck} C.,  et~al., 2019, \mn@doi [\aap] {10.1051/0004-6361/201936169},
  \href {https://ui.adsabs.harvard.edu/abs/2019A&A...631A.167D} {631, A167}

\bibitem[\protect\citeauthoryear{{De Looze} et~al.,}{{De Looze}
  et~al.}{2014}]{DELOOZE14}
{De Looze} I.,  et~al., 2014, \mn@doi [\aap] {10.1051/0004-6361/201322489},
  \href {http://adsabs.harvard.edu/abs/2014A%26A...568A..62D} {568, A62}

\bibitem[\protect\citeauthoryear{{Eldridge}, {Stanway}, {Xiao}, {McClelland},
  {Taylor}, {Ng}, {Greis}  \& {Bray}}{{Eldridge} et~al.}{2017}]{ELDRIDGE17}
{Eldridge} J.~J.,  {Stanway} E.~R.,  {Xiao} L.,  {McClelland} L.~A.~S.,
  {Taylor} G.,  {Ng} M.,  {Greis} S.~M.~L.,   {Bray} J.~C.,  2017, \mn@doi
  [\pasa] {10.1017/pasa.2017.51}, \href
  {https://ui.adsabs.harvard.edu/abs/2017PASA...34...58E} {34, e058}

\bibitem[\protect\citeauthoryear{{Faisst} et~al.,}{{Faisst}
  et~al.}{2016a}]{FAISST16a}
{Faisst} A.~L.,  et~al., 2016a, \mn@doi [\apj] {10.3847/0004-637X/821/2/122},
  \href {http://adsabs.harvard.edu/abs/2016ApJ...821..122F} {821, 122}

\bibitem[\protect\citeauthoryear{{Faisst} et~al.,}{{Faisst}
  et~al.}{2016b}]{FAISST16b}
{Faisst} A.~L.,  et~al., 2016b, \mn@doi [\apj] {10.3847/0004-637X/822/1/29},
  \href {http://adsabs.harvard.edu/abs/2016ApJ...822...29F} {822, 29}

\bibitem[\protect\citeauthoryear{{Faisst} et~al.,}{{Faisst}
  et~al.}{2017}]{FAISST17b}
{Faisst} A.~L.,  et~al., 2017, \mn@doi [\apj] {10.3847/1538-4357/aa886c}, \href
  {http://adsabs.harvard.edu/abs/2017ApJ...847...21F} {847, 21}

\bibitem[\protect\citeauthoryear{{Faisst}, {Capak}, {Emami}, {Tacchella}  \&
  {Larson}}{{Faisst} et~al.}{2019}]{FAISST19b}
{Faisst} A.~L.,  {Capak} P.~L.,  {Emami} N.,  {Tacchella} S.,   {Larson} K.~L.,
   2019, \mn@doi [\apj] {10.3847/1538-4357/ab425b}, \href
  {https://ui.adsabs.harvard.edu/abs/2019ApJ...884..133F} {884, 133}

\bibitem[\protect\citeauthoryear{{Faisst} et~al.,}{{Faisst}
  et~al.}{2020a}]{ALPINE_FAISST20}
{Faisst} A.~L.,  et~al., 2020a, \mn@doi [\apjs] {10.3847/1538-4365/ab7ccd},
  \href {https://ui.adsabs.harvard.edu/abs/2020ApJS..247...61F} {247, 61}

\bibitem[\protect\citeauthoryear{{Faisst}, {Fudamoto}, {Oesch}, {Scoville},
  {Riechers}, {Pavesi}  \& {Capak}}{{Faisst} et~al.}{2020b}]{FAISST20b}
{Faisst} A.~L.,  {Fudamoto} Y.,  {Oesch} P.~A.,  {Scoville} N.,  {Riechers}
  D.~A.,  {Pavesi} R.,   {Capak} P.,  2020b, \mn@doi [\mnras]
  {10.1093/mnras/staa2545}, \href
  {https://ui.adsabs.harvard.edu/abs/2020MNRAS.498.4192F} {498, 4192}

\bibitem[\protect\citeauthoryear{{Ferland} et~al.,}{{Ferland}
  et~al.}{2017}]{FERLAND17}
{Ferland} G.~J.,  et~al., 2017, \rmxaa, \href
  {https://ui.adsabs.harvard.edu/abs/2017RMxAA..53..385F} {53, 385}

\bibitem[\protect\citeauthoryear{{Fudamoto} et~al.,}{{Fudamoto}
  et~al.}{2020}]{ALPINE_FUDAMOTO20}
{Fudamoto} Y.,  et~al., 2020, \mn@doi [\aap] {10.1051/0004-6361/202038163},
  \href {https://ui.adsabs.harvard.edu/abs/2020A&A...643A...4F} {643, A4}

\bibitem[\protect\citeauthoryear{{Giacconi} et~al.,}{{Giacconi}
  et~al.}{2002}]{GIACCONI02}
{Giacconi} R.,  et~al., 2002, \mn@doi [\apjs] {10.1086/338927}, \href
  {https://ui.adsabs.harvard.edu/abs/2002ApJS..139..369G} {139, 369}

\bibitem[\protect\citeauthoryear{{Grogin} et~al.,}{{Grogin}
  et~al.}{2011}]{GROGIN11}
{Grogin} N.~A.,  et~al., 2011, \mn@doi [\apjs] {10.1088/0067-0049/197/2/35},
  \href {https://ui.adsabs.harvard.edu/abs/2011ApJS..197...35G} {197, 35}

\bibitem[\protect\citeauthoryear{{Hao}, {Kennicutt}, {Johnson}, {Calzetti},
  {Dale}  \& {Moustakas}}{{Hao} et~al.}{2011}]{HAO11}
{Hao} C.-N.,  {Kennicutt} R.~C.,  {Johnson} B.~D.,  {Calzetti} D.,  {Dale}
  D.~A.,   {Moustakas} J.,  2011, \mn@doi [\apj] {10.1088/0004-637X/741/2/124},
  \href {http://adsabs.harvard.edu/abs/2011ApJ...741..124H} {741, 124}

\bibitem[\protect\citeauthoryear{{Harikane} et~al.,}{{Harikane}
  et~al.}{2020}]{HARIKANE20}
{Harikane} Y.,  et~al., 2020, \mn@doi [\apj] {10.3847/1538-4357/ab94bd}, \href
  {https://ui.adsabs.harvard.edu/abs/2020ApJ...896...93H} {896, 93}

\bibitem[\protect\citeauthoryear{{Hasinger} et~al.,}{{Hasinger}
  et~al.}{2018}]{HASINGER18}
{Hasinger} G.,  et~al., 2018, \mn@doi [\apj] {10.3847/1538-4357/aabacf}, \href
  {https://ui.adsabs.harvard.edu/abs/2018ApJ...858...77H} {858, 77}

\bibitem[\protect\citeauthoryear{{Hu} et~al.,}{{Hu} et~al.}{2019}]{Hu2019}
{Hu} W.,  et~al., 2019, \mn@doi [\apj] {10.3847/1538-4357/ab4cf4}, \href
  {https://ui.adsabs.harvard.edu/abs/2019ApJ...886...90H} {886, 90}

\bibitem[\protect\citeauthoryear{{Hutchison} et~al.,}{{Hutchison}
  et~al.}{2019}]{HUTCHISON19}
{Hutchison} T.~A.,  et~al., 2019, \mn@doi [\apj] {10.3847/1538-4357/ab22a2},
  \href {https://ui.adsabs.harvard.edu/abs/2019ApJ...879...70H} {879, 70}

\bibitem[\protect\citeauthoryear{{Ichikawa} et~al.,}{{Ichikawa}
  et~al.}{2006}]{Ichikawa2006}
{Ichikawa} T.,  et~al., 2006, in {McLean} I.~S.,  {Iye} M.,  eds,  Society of
  Photo-Optical Instrumentation Engineers (SPIE) Conference Series Vol. 6269,
  Society of Photo-Optical Instrumentation Engineers (SPIE) Conference Series.
  p. 626916, \mn@doi{10.1117/12.670078}

\bibitem[\protect\citeauthoryear{{Ilbert} et~al.,}{{Ilbert}
  et~al.}{2006}]{ILBERT06}
{Ilbert} O.,  et~al., 2006, \mn@doi [\aap] {10.1051/0004-6361:20065138}, \href
  {http://adsabs.harvard.edu/abs/2006A%26A...457..841I} {457, 841}

\bibitem[\protect\citeauthoryear{{Jones}, {Sanders}, {Roberts-Borsani},
  {Ellis}, {Laporte}, {Treu}  \& {Harikane}}{{Jones}
  et~al.}{2020}]{JONESTUCKER20}
{Jones} T.,  {Sanders} R.,  {Roberts-Borsani} G.,  {Ellis} R.~S.,  {Laporte}
  N.,  {Treu} T.,   {Harikane} Y.,  2020, \mn@doi [\apj]
  {10.3847/1538-4357/abb943}, \href
  {https://ui.adsabs.harvard.edu/abs/2020ApJ...903..150J} {903, 150}

\bibitem[\protect\citeauthoryear{{Kaasinen}, {Bian}, {Groves}, {Kewley}  \&
  {Gupta}}{{Kaasinen} et~al.}{2017}]{KAASINEN17}
{Kaasinen} M.,  {Bian} F.,  {Groves} B.,  {Kewley} L.~J.,   {Gupta} A.,  2017,
  \mn@doi [\mnras] {10.1093/mnras/stw2827}, \href
  {https://ui.adsabs.harvard.edu/abs/2017MNRAS.465.3220K} {465, 3220}

\bibitem[\protect\citeauthoryear{{Kashino} et~al.,}{{Kashino}
  et~al.}{2017}]{KASHINO17}
{Kashino} D.,  et~al., 2017, \mn@doi [\apj] {10.3847/1538-4357/835/1/88}, \href
  {http://adsabs.harvard.edu/abs/2017ApJ...835...88K} {835, 88}

\bibitem[\protect\citeauthoryear{{Kennicutt}}{{Kennicutt}}{1998}]{KENNICUTT98}
{Kennicutt} Jr. R.~C.,  1998, \mn@doi [\araa] {10.1146/annurev.astro.36.1.189},
  \href {http://adsabs.harvard.edu/abs/1998ARA%26A..36..189K} {36, 189}

\bibitem[\protect\citeauthoryear{{Kennicutt} \& {De Los Reyes}}{{Kennicutt} \&
  {De Los Reyes}}{2021}]{KENNICUTT21}
{Kennicutt} Robert~C. J.,  {De Los Reyes} M. A.~C.,  2021, \mn@doi [\apj]
  {10.3847/1538-4357/abd3a2}, \href
  {https://ui.adsabs.harvard.edu/abs/2021ApJ...908...61K} {908, 61}

\bibitem[\protect\citeauthoryear{{Kewley}, {Geller}  \& {Jansen}}{{Kewley}
  et~al.}{2004}]{KEWLEY04}
{Kewley} L.~J.,  {Geller} M.~J.,   {Jansen} R.~A.,  2004, \mn@doi [\aj]
  {10.1086/382723}, \href
  {https://ui.adsabs.harvard.edu/abs/2004AJ....127.2002K} {127, 2002}

\bibitem[\protect\citeauthoryear{{Khusanova} et~al.,}{{Khusanova}
  et~al.}{2020}]{ALPINE_KHUSANOVA20}
{Khusanova} Y.,  et~al., 2020, \mn@doi [\aap] {10.1051/0004-6361/201935400},
  \href {https://ui.adsabs.harvard.edu/abs/2020A&A...634A..97K} {634, A97}

\bibitem[\protect\citeauthoryear{{Koekemoer} et~al.,}{{Koekemoer}
  et~al.}{2007}]{KOEKEMOER07}
{Koekemoer} A.~M.,  et~al., 2007, \mn@doi [\apjs] {10.1086/520086}, \href
  {https://ui.adsabs.harvard.edu/abs/2007ApJS..172..196K} {172, 196}

\bibitem[\protect\citeauthoryear{{Kong}, {Charlot}, {Brinchmann}  \&
  {Fall}}{{Kong} et~al.}{2004}]{KONG04}
{Kong} X.,  {Charlot} S.,  {Brinchmann} J.,   {Fall} S.~M.,  2004, \mn@doi
  [\mnras] {10.1111/j.1365-2966.2004.07556.x}, \href
  {http://adsabs.harvard.edu/abs/2004MNRAS.349..769K} {349, 769}

\bibitem[\protect\citeauthoryear{{Kriek} et~al.,}{{Kriek}
  et~al.}{2015}]{KRIEK15}
{Kriek} M.,  et~al., 2015, \mn@doi [\apjs] {10.1088/0067-0049/218/2/15}, \href
  {https://ui.adsabs.harvard.edu/abs/2015ApJS..218...15K} {218, 15}

\bibitem[\protect\citeauthoryear{{Laigle} et~al.,}{{Laigle}
  et~al.}{2016}]{LAIGLE16}
{Laigle} C.,  et~al., 2016, \mn@doi [\apjs] {10.3847/0067-0049/224/2/24}, \href
  {http://adsabs.harvard.edu/abs/2016ApJS..224...24L} {224, 24}

\bibitem[\protect\citeauthoryear{{Le F{\`e}vre} et~al.,}{{Le F{\`e}vre}
  et~al.}{2015}]{LEFEVRE15}
{Le F{\`e}vre} O.,  et~al., 2015, \mn@doi [\aap] {10.1051/0004-6361/201423829},
  \href {http://adsabs.harvard.edu/abs/2015A%26A...576A..79L} {576, A79}

\bibitem[\protect\citeauthoryear{{Le F{\`e}vre} et~al.,}{{Le F{\`e}vre}
  et~al.}{2020}]{ALPINE_LEFEVRE20}
{Le F{\`e}vre} O.,  et~al., 2020, \mn@doi [\aap] {10.1051/0004-6361/201936965},
  \href {https://ui.adsabs.harvard.edu/abs/2020A&A...643A...1L} {643, A1}

\bibitem[\protect\citeauthoryear{{Leitherer}, {Tremonti}, {Heckman}  \&
  {Calzetti}}{{Leitherer} et~al.}{2011}]{LEITHERER11}
{Leitherer} C.,  {Tremonti} C.~A.,  {Heckman} T.~M.,   {Calzetti} D.,  2011,
  \mn@doi [\aj] {10.1088/0004-6256/141/2/37}, \href
  {https://ui.adsabs.harvard.edu/abs/2011AJ....141...37L} {141, 37}

\bibitem[\protect\citeauthoryear{{Lemaux} et~al.,}{{Lemaux}
  et~al.}{2018}]{Lemaux2018}
{Lemaux} B.~C.,  et~al., 2018, \mn@doi [\aap] {10.1051/0004-6361/201730870},
  \href {https://ui.adsabs.harvard.edu/abs/2018A&A...615A..77L} {615, A77}

\bibitem[\protect\citeauthoryear{{Lemaux} et~al.,}{{Lemaux}
  et~al.}{2020}]{Lemaux20}
{Lemaux} B.~C.,  et~al., 2020, arXiv e-prints, \href
  {https://ui.adsabs.harvard.edu/abs/2020arXiv200903324L} {p. arXiv:2009.03324}

\bibitem[\protect\citeauthoryear{{Lubin}, {Gal}, {Lemaux}, {Kocevski}  \&
  {Squires}}{{Lubin} et~al.}{2009}]{Lubin2009}
{Lubin} L.~M.,  {Gal} R.~R.,  {Lemaux} B.~C.,  {Kocevski} D.~D.,   {Squires}
  G.~K.,  2009, \mn@doi [\aj] {10.1088/0004-6256/137/6/4867}, \href
  {https://ui.adsabs.harvard.edu/abs/2009AJ....137.4867L} {137, 4867}

\bibitem[\protect\citeauthoryear{{Maiolino} \& {Mannucci}}{{Maiolino} \&
  {Mannucci}}{2019}]{MAIOLINO19}
{Maiolino} R.,  {Mannucci} F.,  2019, \mn@doi [\aapr]
  {10.1007/s00159-018-0112-2}, \href
  {https://ui.adsabs.harvard.edu/abs/2019A&ARv..27....3M} {27, 3}

\bibitem[\protect\citeauthoryear{{M{\'a}rmol-Queralt{\'o}}, {McLure}, {Cullen},
  {Dunlop}, {Fontana}  \& {McLeod}}{{M{\'a}rmol-Queralt{\'o}}
  et~al.}{2016}]{MARMOLQUERALTO16}
{M{\'a}rmol-Queralt{\'o}} E.,  {McLure} R.~J.,  {Cullen} F.,  {Dunlop} J.~S.,
  {Fontana} A.,   {McLeod} D.~J.,  2016, \mn@doi [\mnras]
  {10.1093/mnras/stw1212}, \href
  {https://ui.adsabs.harvard.edu/abs/2016MNRAS.460.3587M} {460, 3587}

\bibitem[\protect\citeauthoryear{{Masters} et~al.,}{{Masters}
  et~al.}{2014}]{MASTERS14}
{Masters} D.,  et~al., 2014, \mn@doi [\apj] {10.1088/0004-637X/785/2/153},
  \href {http://adsabs.harvard.edu/abs/2014ApJ...785..153M} {785, 153}

\bibitem[\protect\citeauthoryear{{Masters}, {Faisst}  \& {Capak}}{{Masters}
  et~al.}{2016}]{MASTERS16}
{Masters} D.,  {Faisst} A.,   {Capak} P.,  2016, \mn@doi [\apj]
  {10.3847/0004-637X/828/1/18}, \href
  {http://adsabs.harvard.edu/abs/2016ApJ...828...18M} {828, 18}

\bibitem[\protect\citeauthoryear{{McCracken} et~al.,}{{McCracken}
  et~al.}{2012a}]{McCracken2012}
{McCracken} H.~J.,  et~al., 2012a, \mn@doi [\aap]
  {10.1051/0004-6361/201219507}, \href
  {https://ui.adsabs.harvard.edu/abs/2012A&A...544A.156M} {544, A156}

\bibitem[\protect\citeauthoryear{{McCracken} et~al.,}{{McCracken}
  et~al.}{2012b}]{MCCRACKEN12}
{McCracken} H.~J.,  et~al., 2012b, \mn@doi [\aap]
  {10.1051/0004-6361/201219507}, \href
  {http://adsabs.harvard.edu/abs/2012A%26A...544A.156M} {544, A156}

\bibitem[\protect\citeauthoryear{{McLean} et~al.,}{{McLean}
  et~al.}{2010}]{MCLEAN10}
{McLean} I.~S.,  et~al., 2010, in \procspie. p. 77351E,
  \mn@doi{10.1117/12.856715}

\bibitem[\protect\citeauthoryear{{McLean} et~al.,}{{McLean}
  et~al.}{2012}]{MCLEAN12}
{McLean} I.~S.,  et~al., 2012, in \procspie. p. 84460J,
  \mn@doi{10.1117/12.924794}

\bibitem[\protect\citeauthoryear{{Meurer}, {Heckman}  \& {Calzetti}}{{Meurer}
  et~al.}{1999}]{MEURER99}
{Meurer} G.~R.,  {Heckman} T.~M.,   {Calzetti} D.,  1999, \mn@doi [\apj]
  {10.1086/307523}, \href {http://adsabs.harvard.edu/abs/1999ApJ...521...64M}
  {521, 64}

\bibitem[\protect\citeauthoryear{{Nakajima} \& {Ouchi}}{{Nakajima} \&
  {Ouchi}}{2014}]{NAKAJIMA14}
{Nakajima} K.,  {Ouchi} M.,  2014, \mn@doi [\mnras] {10.1093/mnras/stu902},
  \href {http://adsabs.harvard.edu/abs/2014MNRAS.442..900N} {442, 900}

\bibitem[\protect\citeauthoryear{{Nicholls}, {Sutherland}, {Dopita}, {Kewley}
  \& {Groves}}{{Nicholls} et~al.}{2017}]{NICHOLLS17}
{Nicholls} D.~C.,  {Sutherland} R.~S.,  {Dopita} M.~A.,  {Kewley} L.~J.,
  {Groves} B.~A.,  2017, \mn@doi [\mnras] {10.1093/mnras/stw3235}, \href
  {https://ui.adsabs.harvard.edu/abs/2017MNRAS.466.4403N} {466, 4403}

\bibitem[\protect\citeauthoryear{{Nieva} \& {Przybilla}}{{Nieva} \&
  {Przybilla}}{2012}]{NIEVA12}
{Nieva} M.~F.,  {Przybilla} N.,  2012, \mn@doi [\aap]
  {10.1051/0004-6361/201118158}, \href
  {https://ui.adsabs.harvard.edu/abs/2012A&A...539A.143N} {539, A143}

\bibitem[\protect\citeauthoryear{{Noll}, {Burgarella}, {Giovannoli}, {Buat},
  {Marcillac}  \& {Mu{\~n}oz-Mateos}}{{Noll} et~al.}{2009}]{NOLL09}
{Noll} S.,  {Burgarella} D.,  {Giovannoli} E.,  {Buat} V.,  {Marcillac} D.,
  {Mu{\~n}oz-Mateos} J.~C.,  2009, \mn@doi [\aap]
  {10.1051/0004-6361/200912497}, \href
  {https://ui.adsabs.harvard.edu/abs/2009A&A...507.1793N} {507, 1793}

\bibitem[\protect\citeauthoryear{{Oke}}{{Oke}}{1974}]{OKE74}
{Oke} J.~B.,  1974, \mn@doi [\apjs] {10.1086/190287}, \href
  {http://adsabs.harvard.edu/abs/1974ApJS...27...21O} {27, 21}

\bibitem[\protect\citeauthoryear{{Osterbrock}}{{Osterbrock}}{1974}]{OSTERBROCK74}
{Osterbrock} D.~E.,  1974, {Astrophysics of gaseous nebulae}

\bibitem[\protect\citeauthoryear{{Pavesi}, {Riechers}, {Faisst}, {Stacey}  \&
  {Capak}}{{Pavesi} et~al.}{2019}]{PAVESI19}
{Pavesi} R.,  {Riechers} D.~A.,  {Faisst} A.~L.,  {Stacey} G.~J.,   {Capak}
  P.~L.,  2019, \mn@doi [\apj] {10.3847/1538-4357/ab3a46}, \href
  {https://ui.adsabs.harvard.edu/abs/2019ApJ...882..168P} {882, 168}

\bibitem[\protect\citeauthoryear{{Pelliccia} et~al.,}{{Pelliccia}
  et~al.}{2021}]{pelliccia2021}
{Pelliccia} D.,  et~al., 2021, \mn@doi [\apjl] {10.3847/2041-8213/abdf56},
  \href {https://ui.adsabs.harvard.edu/abs/2021ApJ...908L..30P} {908, L30}

\bibitem[\protect\citeauthoryear{{Pettini} \& {Pagel}}{{Pettini} \&
  {Pagel}}{2004}]{PETTINI04}
{Pettini} M.,  {Pagel} B.~E.~J.,  2004, \mn@doi [\mnras]
  {10.1111/j.1365-2966.2004.07591.x}, \href
  {http://adsabs.harvard.edu/abs/2004MNRAS.348L..59P} {348, L59}

\bibitem[\protect\citeauthoryear{{Pickles}}{{Pickles}}{1998}]{PICKLES98}
{Pickles} A.~J.,  1998, \mn@doi [\pasp] {10.1086/316197}, \href
  {https://ui.adsabs.harvard.edu/abs/1998PASP..110..863P} {110, 863}

\bibitem[\protect\citeauthoryear{{Rasappu}, {Smit}, {Labb{\'e}}, {Bouwens},
  {Stark}, {Ellis}  \& {Oesch}}{{Rasappu} et~al.}{2016}]{RASAPPU16}
{Rasappu} N.,  {Smit} R.,  {Labb{\'e}} I.,  {Bouwens} R.~J.,  {Stark} D.~P.,
  {Ellis} R.~S.,   {Oesch} P.~A.,  2016, \mn@doi [\mnras]
  {10.1093/mnras/stw1484}, \href
  {http://adsabs.harvard.edu/abs/2016MNRAS.461.3886R} {461, 3886}

\bibitem[\protect\citeauthoryear{{Reddy} et~al.,}{{Reddy}
  et~al.}{2015}]{REDDY15}
{Reddy} N.~A.,  et~al., 2015, \mn@doi [\apj] {10.1088/0004-637X/806/2/259},
  \href {http://adsabs.harvard.edu/abs/2015ApJ...806..259R} {806, 259}

\bibitem[\protect\citeauthoryear{{Rodriguez-Munoz et al.}}{{Rodriguez-Munoz et
  al.}}{prep}]{RODRIGUEZMUNOZ21}
{Rodriguez-Munoz et al.} L.,  2021 in prep, \mnras

\bibitem[\protect\citeauthoryear{{Saito}, {de la Torre}, {Ilbert}, {Dubois},
  {Yabe}  \& {Coupon}}{{Saito} et~al.}{2020}]{SAITO20}
{Saito} S.,  {de la Torre} S.,  {Ilbert} O.,  {Dubois} C.,  {Yabe} K.,
  {Coupon} J.,  2020, \mn@doi [\mnras] {10.1093/mnras/staa727}, \href
  {https://ui.adsabs.harvard.edu/abs/2020MNRAS.494..199S} {494, 199}

\bibitem[\protect\citeauthoryear{{Salpeter}}{{Salpeter}}{1955}]{SALPETER55}
{Salpeter} E.~E.,  1955, \mn@doi [\apj] {10.1086/145971}, \href
  {http://adsabs.harvard.edu/abs/1955ApJ...121..161S} {121, 161}

\bibitem[\protect\citeauthoryear{{Sanders} et~al.,}{{Sanders}
  et~al.}{2016}]{SANDERS16}
{Sanders} R.~L.,  et~al., 2016, \mn@doi [\apjl] {10.3847/2041-8205/825/2/L23},
  \href {http://adsabs.harvard.edu/abs/2016ApJ...825L..23S} {825, L23}

\bibitem[\protect\citeauthoryear{{Sanders} et~al.,}{{Sanders}
  et~al.}{2020}]{SANDERS20}
{Sanders} R.~L.,  et~al., 2020, \mn@doi [\mnras] {10.1093/mnras/stz3032}, \href
  {https://ui.adsabs.harvard.edu/abs/2020MNRAS.491.1427S} {491, 1427}

\bibitem[\protect\citeauthoryear{{Schaerer} et~al.,}{{Schaerer}
  et~al.}{2020}]{ALPINE_SCHAERER20}
{Schaerer} D.,  et~al., 2020, \mn@doi [\aap] {10.1051/0004-6361/202037617},
  \href {https://ui.adsabs.harvard.edu/abs/2020A&A...643A...3S} {643, A3}

\bibitem[\protect\citeauthoryear{{Schreiber} et~al.,}{{Schreiber}
  et~al.}{2015}]{SCHREIBER15}
{Schreiber} C.,  et~al., 2015, \mn@doi [\aap] {10.1051/0004-6361/201425017},
  \href {http://adsabs.harvard.edu/abs/2015A%26A...575A..74S} {575, A74}

\bibitem[\protect\citeauthoryear{{Scoville} et~al.,}{{Scoville}
  et~al.}{2007}]{SCOVILLE07}
{Scoville} N.,  et~al., 2007, \mn@doi [\apjs] {10.1086/516585}, \href
  {http://adsabs.harvard.edu/abs/2007ApJS..172....1S} {172, 1}

\bibitem[\protect\citeauthoryear{{Shapley} et~al.,}{{Shapley}
  et~al.}{2017}]{SHAPLEY17}
{Shapley} A.~E.,  et~al., 2017, \mn@doi [\apjl] {10.3847/2041-8213/aa8815},
  \href {https://ui.adsabs.harvard.edu/abs/2017ApJ...846L..30S} {846, L30}

\bibitem[\protect\citeauthoryear{{Shen} et~al.,}{{Shen} et~al.}{2021}]{Shen21}
{Shen} L.,  et~al., 2021, \mn@doi [\apj] {10.3847/1538-4357/abee75}, \href
  {https://ui.adsabs.harvard.edu/abs/2021ApJ...912...60S} {912, 60}

\bibitem[\protect\citeauthoryear{{Shim}, {Chary}, {Dickinson}, {Lin},
  {Spinrad}, {Stern}  \& {Yan}}{{Shim} et~al.}{2011}]{SHIM11}
{Shim} H.,  {Chary} R.-R.,  {Dickinson} M.,  {Lin} L.,  {Spinrad} H.,  {Stern}
  D.,   {Yan} C.-H.,  2011, \mn@doi [\apj] {10.1088/0004-637X/738/1/69}, \href
  {http://adsabs.harvard.edu/abs/2011ApJ...738...69S} {738, 69}

\bibitem[\protect\citeauthoryear{{Shimakawa} et~al.,}{{Shimakawa}
  et~al.}{2015}]{SHIMAKAWA15}
{Shimakawa} R.,  et~al., 2015, \mn@doi [\mnras] {10.1093/mnras/stv915}, \href
  {https://ui.adsabs.harvard.edu/abs/2015MNRAS.451.1284S} {451, 1284}

\bibitem[\protect\citeauthoryear{{Speagle}, {Steinhardt}, {Capak}  \&
  {Silverman}}{{Speagle} et~al.}{2014}]{SPEAGLE14}
{Speagle} J.~S.,  {Steinhardt} C.~L.,  {Capak} P.~L.,   {Silverman} J.~D.,
  2014, \mn@doi [\apjs] {10.1088/0067-0049/214/2/15}, \href
  {http://adsabs.harvard.edu/abs/2014ApJS..214...15S} {214, 15}

\bibitem[\protect\citeauthoryear{{Stanway} \& {Eldridge}}{{Stanway} \&
  {Eldridge}}{2018}]{STANWAY18}
{Stanway} E.~R.,  {Eldridge} J.~J.,  2018, \mn@doi [\mnras]
  {10.1093/mnras/sty1353}, \href
  {https://ui.adsabs.harvard.edu/abs/2018MNRAS.479...75S} {479, 75}

\bibitem[\protect\citeauthoryear{{Stark}, {Schenker}, {Ellis}, {Robertson},
  {McLure}  \& {Dunlop}}{{Stark} et~al.}{2013}]{STARK13}
{Stark} D.~P.,  {Schenker} M.~A.,  {Ellis} R.,  {Robertson} B.,  {McLure} R.,
  {Dunlop} J.,  2013, \mn@doi [\apj] {10.1088/0004-637X/763/2/129}, \href
  {http://adsabs.harvard.edu/abs/2013ApJ...763..129S} {763, 129}

\bibitem[\protect\citeauthoryear{{Stark} et~al.,}{{Stark}
  et~al.}{2014}]{STARK14}
{Stark} D.~P.,  et~al., 2014, \mn@doi [\mnras] {10.1093/mnras/stu1618}, \href
  {https://ui.adsabs.harvard.edu/abs/2014MNRAS.445.3200S} {445, 3200}

\bibitem[\protect\citeauthoryear{{Stark} et~al.,}{{Stark}
  et~al.}{2015}]{STARK15}
{Stark} D.~P.,  et~al., 2015, \mn@doi [\mnras] {10.1093/mnras/stv1907}, \href
  {https://ui.adsabs.harvard.edu/abs/2015MNRAS.454.1393S} {454, 1393}

\bibitem[\protect\citeauthoryear{{Steidel} et~al.,}{{Steidel}
  et~al.}{2014}]{STEIDEL14}
{Steidel} C.~C.,  et~al., 2014, \mn@doi [\apj] {10.1088/0004-637X/795/2/165},
  \href {http://adsabs.harvard.edu/abs/2014ApJ...795..165S} {795, 165}

\bibitem[\protect\citeauthoryear{{Strom}, {Steidel}, {Rudie}, {Trainor},
  {Pettini}  \& {Reddy}}{{Strom} et~al.}{2017}]{STROM17}
{Strom} A.~L.,  {Steidel} C.~C.,  {Rudie} G.~C.,  {Trainor} R.~F.,  {Pettini}
  M.,   {Reddy} N.~A.,  2017, \mn@doi [\apj] {10.3847/1538-4357/836/2/164},
  \href {http://adsabs.harvard.edu/abs/2017ApJ...836..164S} {836, 164}

\bibitem[\protect\citeauthoryear{{Suzuki} et~al.,}{{Suzuki}
  et~al.}{2008}]{Suzuki2008}
{Suzuki} R.,  et~al., 2008, \mn@doi [\pasj] {10.1093/pasj/60.6.1347}, \href
  {https://ui.adsabs.harvard.edu/abs/2008PASJ...60.1347S} {60, 1347}

\bibitem[\protect\citeauthoryear{{Swinbank}, {Bower}, {Smith}, {Wilman},
  {Smail}, {Ellis}, {Morris}  \& {Kneib}}{{Swinbank} et~al.}{2007}]{SWINBANK07}
{Swinbank} A.~M.,  {Bower} R.~G.,  {Smith} G.~P.,  {Wilman} R.~J.,  {Smail} I.,
   {Ellis} R.~S.,  {Morris} S.~L.,   {Kneib} J.~P.,  2007, \mn@doi [\mnras]
  {10.1111/j.1365-2966.2007.11454.x}, \href
  {https://ui.adsabs.harvard.edu/abs/2007MNRAS.376..479S} {376, 479}

\bibitem[\protect\citeauthoryear{{Swinbank} et~al.,}{{Swinbank}
  et~al.}{2009}]{SWINBANK09}
{Swinbank} A.~M.,  et~al., 2009, \mn@doi [\mnras]
  {10.1111/j.1365-2966.2009.15617.x}, \href
  {https://ui.adsabs.harvard.edu/abs/2009MNRAS.400.1121S} {400, 1121}

\bibitem[\protect\citeauthoryear{{Tanaka} et~al.,}{{Tanaka}
  et~al.}{2011}]{Tanaka2011}
{Tanaka} I.,  et~al., 2011, \mn@doi [\pasj] {10.1093/pasj/63.sp2.S415}, \href
  {https://ui.adsabs.harvard.edu/abs/2011PASJ...63S.415T} {63, 415}

\bibitem[\protect\citeauthoryear{{Troncoso} et~al.,}{{Troncoso}
  et~al.}{2014}]{TRONCOSO14}
{Troncoso} P.,  et~al., 2014, \mn@doi [\aap] {10.1051/0004-6361/201322099},
  \href {https://ui.adsabs.harvard.edu/abs/2014A&A...563A..58T} {563, A58}

\bibitem[\protect\citeauthoryear{{Vallini}, {Ferrara}, {Pallottini}, {Carniani}
   \& {Gallerani}}{{Vallini} et~al.}{2021}]{VALLINI21}
{Vallini} L.,  {Ferrara} A.,  {Pallottini} A.,  {Carniani} S.,   {Gallerani}
  S.,  2021, \mn@doi [\mnras] {10.1093/mnras/stab1674}, \href
  {https://ui.adsabs.harvard.edu/abs/2021MNRAS.505.5543V} {505, 5543}

\bibitem[\protect\citeauthoryear{{Yang} \& {Lidz}}{{Yang} \&
  {Lidz}}{2020}]{YANGLIDZ20}
{Yang} S.,  {Lidz} A.,  2020, \mn@doi [\mnras] {10.1093/mnras/staa3000}, \href
  {https://ui.adsabs.harvard.edu/abs/2020MNRAS.499.3417Y} {499, 3417}

\bibitem[\protect\citeauthoryear{{Yeh} \& {Matzner}}{{Yeh} \&
  {Matzner}}{2012}]{YEH12}
{Yeh} S. C.~C.,  {Matzner} C.~D.,  2012, \mn@doi [\apj]
  {10.1088/0004-637X/757/2/108}, \href
  {https://ui.adsabs.harvard.edu/abs/2012ApJ...757..108Y} {757, 108}

\makeatother
\end{thebibliography}




\appendix

\section{Absolute Flux Calibration of the \textit{MOSFIRE} Spectrum} \label{sec:absolutecalibration}

For the absolute calibration of the \textit{MOSFIRE} spectrum, we use a $17^{\rm th}$ magnitude standard star from the 2MASS star catalogue, which was included for this purpose in the mask. The one-dimensional spectrum of the star is extracted within $10\,{\rm pixels}$ ($1.8\arcsec$) around the centre of the continuum emission on the two-dimensional spectrum in ${\rm e^{-}/s}$. The wavelength-dependent calibration is then derived by comparing the observed spectrum with models of different stellar spectral types.

\begin{figure}
    \centering
    \includegraphics[width=0.48\textwidth]{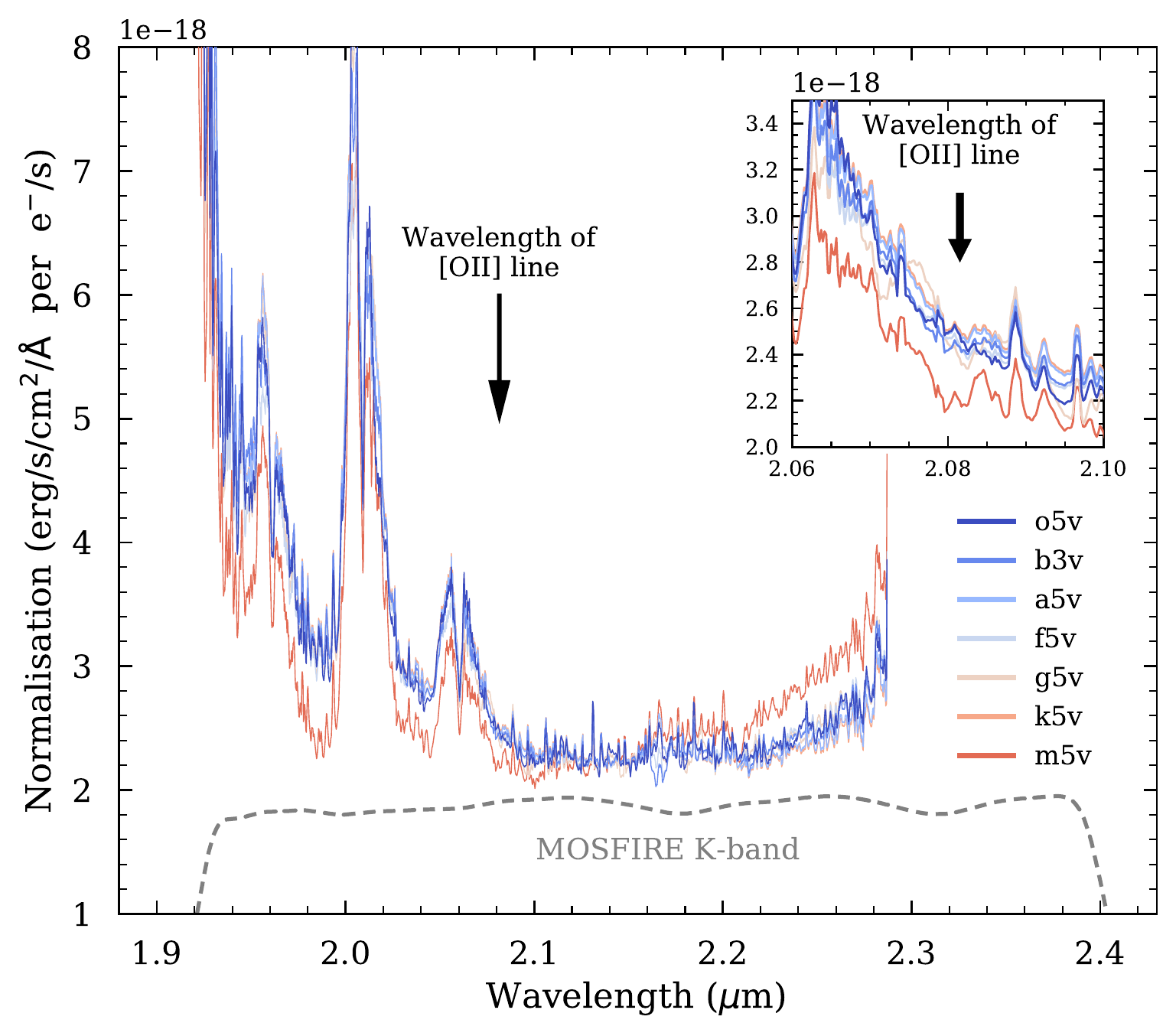}
    \caption{Wavelength-dependent normalisation of the \textit{MOSFIRE}/K band spectrum derived for various stellar types from the \citet[][]{PICKLES98} library. Our standard star is close to a G5 V type. The wavelength of \oii~for \thegalaxy~is indicated by an arrow. The dashed line represents the wavelength range of the \textit{MOSFIRE} K-band.
    }
    \label{fig:normalization}
\end{figure}

In detail, we use the spectra from the \citet[][]{PICKLES98} stellar model library, which offers a wide range in spectral types from O to M and covers a wavelength from $0.115-2.5\,{\rm \mu m}$ in steps of $5\,{\rm \A}$. This is sufficiently red to match the \textit{MOSFIRE} $K-$band. We find that the spectrum at $>2\,{\rm \mu m}$ only changes slowly as a function of stellar type, hence we only focus on a coarse grid including O5 V, B3 V, A5 V, F5 V, G5 V, K5 V, and M5 V type stars. 
The model spectrum of each of these is normalised to the UltraVista $K-$band flux of the observed 2MASS star, obtained from the \textit{COSMOS2015} catalogue \citep[][]{LAIGLE16,MCCRACKEN12}. A wavelength-dependent conversion from units of $\rm e^{-}\,s^{-1}$ to $\rm erg\,s^{-1}\,cm^{-2}\,\A^{-1}$ is then derived by comparing the normalised model spectrum to the observed spectrum of the 2MASS star. The resulting normalisations are shown in Figure~\ref{fig:normalization}. The arrow indicates the wavelength of \oii. The computed normalization is within $5\%$ for all spectral types except the coolest M-dwarf. By comparing the $K-$band normalised stellar spectra to the UltraVISTA $Y$, $J$, and $H$ photometry, we find that our standard star fits best a G$-$type star. We therefore use the G5 V normalisation through out the paper.

\section{Comparison of Optical Emission Line Measurements to \textit{EL-COSMOS}} \label{sec:elcosmos}

\begin{figure}
    \centering
    \includegraphics[width=0.48\textwidth]{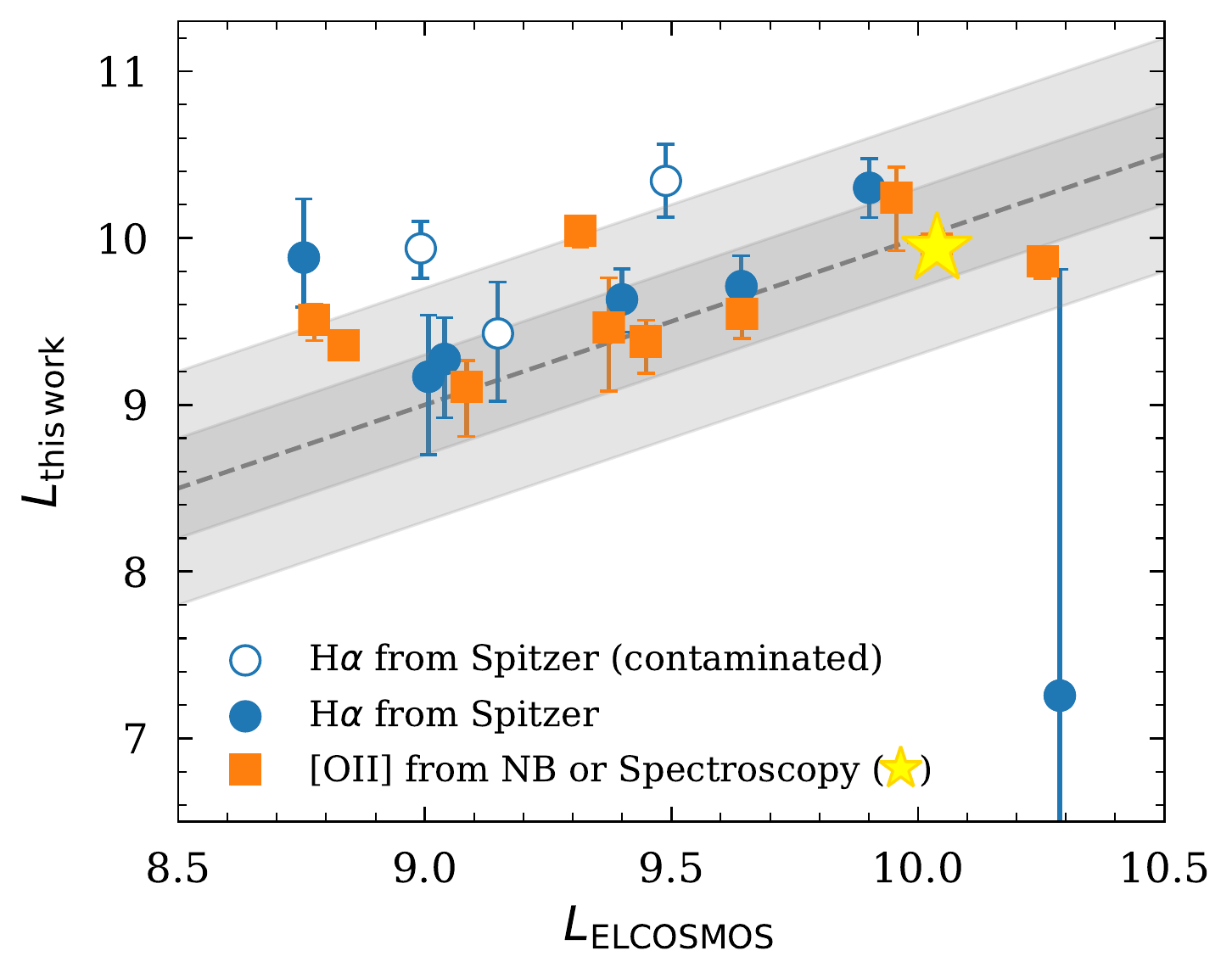}
    \caption{Comparison between our measurements of \halpha~(blue circles, derived from \textit{Spitzer} colours) and \oii~(orange squares, derived from narrow-band imaging or spectroscopy) and the derivations in the \textit{EL-COSMOS} catalog. Galaxies whose \halpha~measurements are uncertain due to contaminated \textit{Spitzer} photometry are shown as open circles. The dark (light) grey areas denote deviations of a factor of $2$ ($5$) from the 1$-$to$-$1 relation (dashed line).
    }
    \label{fig:elcosmos}
\end{figure}

We derived the \halpha~luminosities for our $10$ galaxies from their \textit{Spitzer} \iracA~colours (see \citet[][]{ALPINE_FAISST20} and \citet[][]{FAISST16a} for a detailed description of the methods used). Here we compare our measurement to the recent \textit{EL-COSMOS} catalogue \citep[][]{SAITO20}, which provides predictions of the intrinsic (dust-corrected) \oii~and \halpha~optical lines from SED fitting calibrated to spectroscopic measurements for all galaxies in the \citet[][]{LAIGLE16} \textit{COSMOS2015} catalogue.
Our luminosity measurements have been dust corrected using the method described in the this paper.
Figure~\ref{fig:elcosmos} shows the result of the comparison. \textit{EL-COSMOS} agrees very well with our narrow-band and spectroscopic \oii~measurements. Comparing the \halpha~measurements, we find a good agreement within a factor of $2$. Note that we marked galaxies that have uncertain \halpha~measurements due to contaminated \textit{Spitzer} photometry \citep[see also][]{ALPINE_FAISST20}.

\bsp	
\label{lastpage}
\end{document}